\documentclass[sn-mathphys,Numbered]{sn-jnl}

\usepackage{graphicx}%
\usepackage{multirow}%
\usepackage{amsmath,amssymb,amsfonts}%
\usepackage{mathrsfs}%
\usepackage[title]{appendix}%
\usepackage{xcolor}%
\usepackage{textcomp}%
\usepackage{manyfoot}%
\usepackage{booktabs}%
\usepackage{algorithm}%
\usepackage{algorithmicx}%
\usepackage{algpseudocode}%
\usepackage{listings}%
\usepackage{hyperref}

\newcommand{\trans}[1]{\stackrel{#1}{\longrightarrow}}
\newcommand{\ntrans}[1]{\stackrel{#1}{\mathrel{\not\hspace{-.4em}\longrightarrow}}}

\begin{document}

\title[Formal Methods for Mobile Ad Hoc Networks: A Survey]{Formal Methods for Mobile Ad Hoc Networks:\\ A Survey}

\author*[1]{\fnm{Wan} \sur{Fokkink}}\email{w.j.fokkink@vu.nl}
\author[2]{\fnm{Rob} \sur{van Glabbeek}}\email{rvg@cs.stanford.edu}

\affil*[1]{\orgdiv{Department of Computer Science}, \orgname{Vrije Universiteit Amsterdam}, \orgaddress{\street{\mbox{De Boelelaan} 1111}, \postcode{1081 HV} \city{Amsterdam}, \country{The Netherlands}}}
\affil[2]{\orgdiv{School of Informatics}, \orgname{University of Edinburgh}, \orgaddress{\street{10 Crichton Street}, \city{Edinburgh}, \postcode{EH8 9AB}, \country{United Kingdom}}}

\abstract{In a mobile ad hoc network (MANET), communication is wireless and nodes can move independently. Properly analyzing the functional correctness, performance, and security of MANET protocols is a challenging task. A wide range of formal specification and analysis techniques have been employed in the analysis of MANET protocols. This survey presents an overview of rigorous formal analysis techniques and their applications, with a focus on MANET routing protocols. Next to functional correctness, also real-time properties and security are considered. Moreover, an overview is given of formal frameworks that target MANETs specifically, as well as mobility models that underlie performance analyses of MANET protocols. The aim is to give a comprehensive and coherent overview of this rather scattered field, in which a variety of rigorous formal methods have been applied to analyze different aspects of a wide range of MANET protocols.}


\maketitle

\section{Introduction}

A mobile ad hoc network, abbreviated to MANET, is a wireless network in which the nodes are free to move independently, so that the network topology changes over time. Its origins date back to the 1970s, when the U.S.\ Defense Research developed the Packet Radio NETwork \cite{JubinTornow87} and its follow-on the SURvivable Adaptive Network \cite{ShachamWestcott87}. Interest in MANETs has grown significantly over the last two decades, owing to the common availability of wireless communication devices, connecting e.g.\ cell phones and laptops. MANETs may dynamically self-organize and self-configure, which facilitates their employment for e.g.\ mobile robots and in dire circumstances such as an earthquake area. They are currently widely used in practice, notably in disaster relief and maritime applications.

Network connectivity in a MANET is typically based on broadcast communication, in which a message is received by all nodes within a certain range of the sender. A key challenge is to let nodes continuously maintain information needed to route data traffic through the network, over multiple hops from the source to the destination of a data packet. This is complicated by the fact that MANETs can grow very large and their nodes often have only restricted resources and need to share limited communication bandwidth.

A wide range of MANET protocols have been developed over the last 25 years. Their functional correctness and performance were mostly analyzed by means of simulations, through for instance the widely employed simulator ns-3. This offers a convenient way to evaluate and compare the performance of MANET protocols in large-scale networks of thousands of nodes and reproduce experiments \cite{DasCY00}. However, adequately simulating wireless mobility is a tall order because important corner cases may easily be missed in the large spectrum of possible behaviors, and performance results tend to be sensitive to networking and user traffic profiles \cite{CavinSS02}. Furthermore, due to their open and dynamic nature, MANETs are vulnerable to attacks by malicious intruders, which makes it extra important to analyze their protocols from a security perspective. In view of these challenges, formal specification and analysis techniques have been applied to analyze MANET protocols. An additional advantage is that formal methods provide precise, unambiguous protocol specifications, compared to the informal textual descriptions of protocol standards.

Different formal modeling frameworks, especially a significant number of process calculi, were developed that target the specification and verification of MANET protocols. In particular, their languages include mechanisms to specify connectivity between network nodes and to let the network topology evolve over time. These frameworks have been applied successfully in the analysis of MANET protocols, in particular for routing. Not only were many of these protocols proven correct formally, also these analyses led to the detection of flaws and security vulnerabilities that induced adaptations in MANET protocol standards.

Coping with the huge number and unpredictability of possible mobility scenarios in a MANET is the key challenge in analyzing its protocols. Therefore it is of the essence to employ mathematical mobility models that are on the one hand comprehensive and at the other hand compact. Different mobility models of nodes in a MANET have been developed and employed in particularly performance analyses of MANET protocols.

The aim of the current survey is to give a comprehensive and coherent overview of this rather scattered field, in which a variety of formal methods have been applied to analyze different aspects of a wide range of MANET protocols, using different mobility models. We discuss analyses of routing protocols for MANETS using existing formal methods, addressing functional correctness as well as real-time and security properties. It is also explained how these analyses have impacted specific protocol standards. We moreover consider several formal frameworks developed specifically for MANETs, with an emphasis on process calculi. Finally, different mobility models are discussed, capturing how nodes move through a network, and explain how such probabilistic models impact performance analyses of MANET protocols.

To arrive at an exhaustive list of research papers in this domain, we took the following steps. First, we performed searches in the DBLP database and Google Scholar with as search terms combinations of firstly formal/model/analysis and secondly MANET or a variant of this acronym (mobile/dynamic/ad hoc network). Furthermore, publication lists of prolific researchers in the field were scrutinized. Next, we performed snowballing by investigating for each selected paper on the one hand its reference list to look for older relevant papers, and on the other its Google Scholar entry to look for later relevant papers that cite the one at hand. An Achilles heel of this methodology is that it focused mostly on titles of papers, which is mitigated by the fact that in Google Scholar also bodies of papers were taken into account. The selected papers were grouped in coherent story lines, that show how the research field has progressed over the past 25 years. We discuss formal analysis techniques and tools developed especially for MANET protocols as well as notable case studies, and provide some directions for future research.

Formal methods are characterized by three ingredients: syntax, semantics, and analysis technique (see e.g.\ \cite{RoggenbachSS22}). Formal methods come with an exactly defined syntax and a rigorous semantics. Analysis techniques that allow to provide a rigid correctness proof, notably theorem proving, or that perform an exhaustive analysis of the state space of model, notably model checking, are in \cite{KulikDLMSTW22} contrasted with what are called lightweight formal methods. Typical examples of the latter category, which do not provide an exhaustive analysis, are static code analysis and model animation. This survey focuses on the first category of rigorous, exhaustive formal methods, whereby we do take into account some techniques that are tilted toward the second category, notably statistical model checking and model-based conformance testing.

It should be kept in mind that formal analysis techniques tend to be applied to abstractions of protocol standards. First, because specifying such a standard in full detail is cumbersome. Second, because it can be infeasible to prove a formal model of the entire standard correct formally. This holds true especially for MANETs, for which the vast range of possible mobility scenarios can give rise to huge state spaces of protocol behavior. The aim of an abstraction is that all key aspects of the standard are included in the formal model, but inevitably an abstraction leads to a loss of information. In some cases a specification of a MANET protocol was proven correct formally, while the protocol later turned out to be flawed, due to the fact that the abstraction employed in the correctness proof lacked vital ingredients.

The formal analysis of MANET networks is a huge challenge, which has given rise to the development of important new concepts and featured many success stories, but is still an ongoing research effort. This survey gives an overview of work done in this area during the last two decades and discusses some challenges that lie ahead.

This survey is organized as follows. Section \ref{sec:analysis} gives an overview of analyses of MANET protocols using existing formal methods, i.e., not tailored to MANETs. This section emphasizes routing protocols, and distinguishes functional correctness, real-time properties, and security aspects. Section \ref{sec:frameworks} presents formal frameworks that were designed specifically for analyzing MANET protocols, with an emphasis on process calculi. Section \ref{sec:mobility} discusses different ways of modeling mobility in a MANET and how these are used in performance analyses of protocols. Finally, Section \ref{sec:conclusions} provides conclusions and possible avenues for future research.

\section{Analysis of Routing Protocols with Existing Formal Methods}
\label{sec:analysis}

Initially, formal verifications of MANETs focused on the specification and analysis of specific network protocols using existing formal methods that are not tailored to MANETs. In particular, the functional correctness, real-time performance, and security of routing protocols has been studied widely, as these constitute key challenges for MANETs. In link-state routing protocols, such as the Optimized Link State Routing (OLSR) protocol \cite{ClausenJacquet03}, nodes continuously gather information on the local network topologies at other nodes to build a view of the global topology, so that at all times routes can be computed locally. By contrast, on-demand protocols, such as the Ad-hoc On-Demand Distance Vector (AODV) routing protocol \cite{PerkinsRoyer99,PerkinsRD03}, build a multi-hop path by letting the source node of this path send exploration messages through the network toward the destination node. These route request messages carry a hop count to keep track of the length of the path, in order to find a shortest path and abort the attempt if the destination is not found within a certain limit. In case an exploration message reaches the destination, route reply messages travel from the destination toward the source via the established path. If a link break is detected by a node, it sends a route error message in the opposite direction of affected routes, so that nodes on such paths cancel these routes.

Initially, especially model checking and theorem proving were promoted to analyze MANET protocols. Model checking allows one to generate the entire state space of a protocol for one particular network topology and automatically check whether certain requirements, formulated in some temporal logic, are satisfied in the entire state space. With interactive theorem one can prove such requirements mathematically for general network topologies, whereby the human verifier and the theorem proving tool work hand in hand.

\subsection{Functional Correctness}

AODV served as case study in a string of papers, in some cases formally proving its correctness, in others uncovering flaws, in particular concerning the formation of cyclic paths called loops. In early works \cite{DasDill02,EnglerMusuvathi04}, case studies on a high-level specification of AODV highlight the strength of at that time novel model checking techniques. The first paper presents a novel predicate abstraction method to turn infinite-state systems into finite-state ones. The second paper propagates the strength of software model checking, where a piece of software is verified instead of an abstract model.

A notable early paper is \cite{BhargavanOG02}, in which both the model checker SPIN \cite{Holzmann97} and the interactive theorem prover HOL \cite{GordonMelham93} are used for the analysis of a preliminary version of the AODV protocol. SPIN is applied to find different scenarios for AODV in which loops are formed. Moreover, loop freedom is proved for a stricter specification of AODV, with extra conditions that avoid these scenarios. The verification with HOL, under the assumption that nodes never delete routes, is based on an abstraction of hop counts and of sequence numbers, which indicate the freshness of routes. Correctness of this abstraction is left as an open question. In \cite[p. 124]{FehnkerGHMPT13} it is shown that a key invariant (i.e., a property that is satisfied in all reachable states) in this correctness proof, in essence originating from \cite{PerkinsRoyer99}, does not hold for AODV.

Formal proofs of loop freedom for an updated version of AODV are presented in \cite{WuXZ13,ZhouYZW09}. In \cite{GlabbeekHTP13} however it is shown that loops can occur in this version of AODV. The reason for this discrepancy is that in \cite{WuXZ13,ZhouYZW09}, route reply messages generated at intermediate nodes on a path are abstracted away.

In a Petri net tokens concurrently travel through a directed graph; in a colored Petri net \cite{Jensen81} tokens are data values. AODV and its successor AODVv2 \cite{PerkinsRD02} (also known as DYMO) are modeled in \cite{XiongMT02} and analyzed in \cite{EspensenKK08} using colored Petri nets. In the latter paper several underspecifications and issues in the AODV2 standard are reported, notably with regard to the content of route request messages. In \cite{BillingtonYuan09} the model from \cite{EspensenKK08} is optimized so that its state space is reduced considerably. Four ambiguities in the AODV specification are revealed in \cite{GlabbeekHTP13}, in particular concerning how to act at the receipt of a route reply or route error message. Possible interpretations of these ambiguities are considered exhaustively, and it is shown that certain interpretations of these ambiguities give rise to unwanted behavior. Experimentation with five existing open source implementations of AODV show that they deal with these ambiguities in different ways. For more than 5000 interpretations it is verified whether they are loop-free, as explained in \cite{FehnkerGHMPT13}, demonstrating how this formal reasoning approach can be adapted conveniently for protocol variants. In \cite{EdenhoferHofner12} AODVv2 is specified in AWN, a process calculus that will be discussed in Section \ref{sec:process-calculi}, and it is shown that some problems of AODV were resolved in AODVv2 but new shortcomings were introduced. Notably, route request messages may be lost in AODV; a fix of this problem causes the loss of both route request and reply messages in AODVv2. The latter protocol is proven to be loop-free in \cite{SaksenaWJ08} using a graph grammar framework, but is later shown to suffer from loops in \cite{NamjoshiTrefler15b,YousefiGK17}. The latter two papers suggest improvements and formally prove loop freedom for these adapted versions. It is worth noting that the specification issues reported in \cite{EspensenKK08,BillingtonYuan09} and the detected loop formations in \cite{BhargavanOG02,GlabbeekHTP13,NamjoshiTrefler15b,YousefiGK17} led to adaptations in the official AODV standard.

Other MANET routing protocols have also been analyzed using model checking, on very small networks to cope with state space explosion. The underlying idea is that flaws in these protocols may already occur with only a few nodes. In \cite{ZakiuddinGWG03} the refinement checker FDR \cite{Lawrence04} is used to analyze the Cluster-Based Routing Protocol (CBRP) \cite{JiangLT98} on networks of five nodes. In \cite{WiblingPP04} the Lightweight Underlay Network Ad hoc Routing (LUNAR) protocol \cite{TschudinGRW04} is analyzed with SPIN for networks of seven nodes, with respect to a few specific mobility scenarios. In \cite{RenesseAghvami04} the Wireless Adaptive Routing Protocol (WARP) \cite{KhengarAghvami01} is analyzed for networks of five nodes. General mobility is taken into account, leading to huge state spaces. Therefore a simplified version of the protocol is considered and so-called bitstate hashing is employed, meaning that roughly $98\%$ of the state space is covered in the analysis. In \cite{Oleshchuk05} it is explained how connection properties within ad hoc wireless sensor networks can be analyzed using SPIN. In \cite{SteeleAndel12} several functional properties of the OLSR protocol, including loop freedom, are verified with SPIN on networks of four nodes.  Byzantine node failures are taken into account, in which case
the verification provides an execution trace to the failure.

Different techniques have been developed to cope with the state space explosion problem inherent to MANET protocols. In \cite{CamaraLF07} the intermediate nodes in a route are lumped together. SPIN is applied to the abstracted state spaces of three routing protocols, the Location Aided Routing (LAR) protocol \cite{KoVaidya00}, the Distance Routing Effect Algorithm for Mobility (DREAM) \cite{BasagniCSW98}, and OLSR, leading to the detection of flaws in all three protocols. A symbolic reachability algorithm for MANET protocols is introduced in \cite{SinghRS09}, based on a constraint language for representing topologies. In \cite{GhassemiFokkink16} the transitions in the state space of a MANET protocol are endowed with topological connectivity information. Model checking is performed, using the mCRL2 toolset \cite{GrooteKSWW11}, with regard to a temporal logic that takes into account connectivity. In a mobility-preserving abstraction of a state space of a MANET protocol from \cite{NanzNN10}, the transitions are labeled by connectivity information, and model checking can be performed with regard to a three-valued temporal logic. In \cite{NamjoshiTrefler15a} it is demonstrated on AODVv2 that it may suffice to prove global invariants for arbitrarily large dynamic networks of similar nodes on a small ``cutoff'' network. In \cite{KojimaNT16} AODV serves as running example for a novel approach to reduce state spaces of MANET routing protocols by considering only the last attempt of a source node trying to establish a route to some destination.

Model-based conformance testing verifies whether execution traces of a system implementation conform to a formal model of the desired system behavior. A methodology for conformance testing of MANET routing protocols is presented in \cite{MaagGC08}, which tries to address the challenge of interpreting the outcomes of test executions against unpredictable underlying topology changes. The notion of self-similarity from \cite{DjouvasGL06} is exploited: nodes on the same paths of forwarded packets are collapsed, to obtain small networks that are representative of the entire network. The approach is used to test an implementation of the Dynamic Source Routing (DSR) protocol \cite{HuMJ07,JohnsonMB01} against a model of this protocol in the Specification and Description Language (SDL) \cite{BelinaHogrefe89}. Many test verdicts unfortunately remain inconclusive, due to the huge number of possible mobility patterns. In \cite{AndresMCMN09} this framework is employed to passively test the OLSR protocol, meaning that input and output events of the implementation under test are observed at run-time without stimulating the implementation. In \cite{LalanneMaag13} this experiment is repeated to showcase the formal tool DataMonitor for passively testing MANET routing protocols.

\subsection{Real-time Properties}

With regard to the formal analysis of real-time properties of MANET protocols, again \cite{BhargavanOG02} is a notable starting point. Next to functional correctness of the Routing Information Protocol (RIP) \cite{Hedrick88}, a sharp real-time upper bound on the computation of paths is proved using SPIN and HOL, under the assumption that the network topology remains stable. In \cite{WiblingPP04} the LUNAR protocol is analyzed on networks of up to five nodes, with respect to some specific mobility scenarios, using the model checker {\sc Uppaal} for real-time properties \cite{LarsenPY97}. In the follow-up paper \cite{WiblingPP05}, by abstracting away some messages that are redundant in establishing a route, this is pushed up to networks with a diameter of eleven hops. In \cite{ChiyangwaKwiatkowska05} an {\sc Uppaal} analysis of AODV shows that in paths with twelve intermediate nodes, a node may give up waiting for a route reply message too quickly. Their suggestion to make this delay dependent on the network diameter has been incorporated in a subsequent version of the AODV standard.

In \cite{ChaudharyFM17} a formal model from \cite{Furlan12} of the Better Approach to Mobile Ad hoc Networks (B.A.T.M.A.N.) routing protocol \cite{NeumannALW08} is considered. Some ambiguities in the protocol standard are resolved and functional correctness is shown using {\sc Uppaal}. An adaptation of this protocol standard is proposed, and it is shown by means of simulations that this adaptation leads to significantly fewer suboptimal routes. In \cite{FehnkerCM18} the latter analysis is confirmed using {\sc Uppaal}, considering dynamic topologies in a $4\times 4$ grid. In \cite{KamaliHKP15} OLSR is analyzed using {\sc Uppaal}. Next to functional properties on stable networks, also route discovery times are analyzed on dynamic networks, for networks of five nodes. In \cite{KamaliPetre15} this work is extended by proposing a new error message for OLSR and showing with {\sc Uppaal} that this halves recovery time in case of a link failure.

A method to alleviate the state space explosion problem for real-time model checking is proposed in \cite{MouradianAuge-Blum14}. Hard delay bounds for MANET protocols are derived by applying model checking for each individual node only and abstracting node interactions to arrival curves which express upper and lower bounds on the number of events that may arrive over a specified time interval, using Sensor Network Calculus \cite{SchmittRoedig05}. This method is applied to the Real-Time X-layer Protocol (RTXP) \cite{MouradianAV14}, a routing protocol with guaranteed bounded end-to-end delays.

In \cite{HammalSBA17} a novel method is proposed to build an efficient peer-to-peer overlay network on top of a MANET, by selecting physically near and fresh peer-to-peer neighbors. Several safety and liveness
properties for this framework, such as absence of deadlock and successful termination of transmissions, are verified using {\sc Uppaal}, allowing to take into account timeouts at the network nodes.

Statistical model checking mixes traditional model checking with simulation. Execution runs of a model checker on the state space of the specified system are monitored with respect to some property, to obtain a statistical estimate on the validity of the property. In \cite{HofnerMcIver13} this approach is advocated for the analysis of real-time properties of MANETs, to combat the state space explosion problem. A comparison is made between the performance of AODV and AODVv2 on networks with static topologies by means of the stochastic timed model checker {\sc Uppaal} SMC \cite{DavidLLMP15}. In these experiments, AODV performs better than AODVv2 on networks of five nodes. Moreover, it is shown that this analysis could scale to networks of up to a hundred nodes. In \cite{DalCorsoMM15} this experiment is repeated on $4\times 3$ toroidal networks with lossy communication, in which case by contrast AODVv2 performs significantly better than AODV. An explanation for these different experimental outcomes is that on very small networks AODVv2 does not benefit from an optimization called path accumulation, meaning that messages accumulate information about the nodes they visit and distribute this information to their recipients. In \cite{KamaliMD18} it is analyzed how an upgrade of AODVv2 to resolve loop formations, called AODVv2-16, impacts performance, on $3\times 3$ grid networks. It is found that the old version performs significantly better, especially in case of a high rate of message loss. In \cite{KamaliFehnker18} it is explained how {\sc Uppaal} SMC can be used in a structured way to model MANET protocols, by building reusable components for such protocols. This framework is applied to analyze the probability of packet delivery for AODV, B.A.T.M.A.N., and OLSR on different network topologies of nine nodes, at different rates of link failures.

\subsection{Security}

A protocol is considered secure if it is robust against malicious attackers. In a MANET, nodes must be able to trust the intermediate nodes in a route. A malicious node should not be able to secretly inject itself into the network or overhear private information, in spite of the dynamic and open nature of MANETs and wireless communication being vulnerable to malicious activity such as eavesdropping. The importance of an in-depth security analysis of MANET protocols is underscored by the wide range of security issues in the OLSR protocol reported in \cite{HerbergClausen10}.

A strand space \cite{ThayerHG99} is a graph structure of events generated by causal interaction. It exploits that security protocols typically exhibit nonbranching behavior, as the participants tend to execute a fixed sequence of events. The tool Athena \cite{Song99} automatically checks security properties of strand spaces using a combination of model checking and theorem proving. In \cite{YangBaras03} the strand space model and Athena are extended to capture the branching behavior of MANET routing protocols and search for insider attacks. As a case study it is shown on a network of four nodes that in AODV an attacker can forge a route reply message.

Ariadne \cite{HuPJ05} intends to be a secure MANET routing protocol. In \cite{AcsBV06} however attacks on this protocol are revealed, in which the malicious intruder redirects routes through nodes under its own control, and an adaptation called endairA (the reverse of Ariadne) is proposed. While in Ariadne a route request message is digitally signed, in endairA all intermediate nodes sign a route reply message. The endairA protocol is proven correct using simulations of a mathematical model of the protocol. In \cite{BurmesterM09} however another attack is exposed on endairA, in which adversarial nodes exploit nonexisting links to transfer data (such as signatures). It is argued that the mathematical framework put forward in \cite{AcsBV06} for analyzing the security of MANET routing protocols needs to be extended to take into account such hidden channels. A posterior analysis of endairA in \cite{BenettiMV10} with the model checker AVISPA \cite{ArmandoBBCCCDHKMMORSTVV05}, which targets security protocols, inadvertently misses the attack uncovered in \cite{BurmesterM09} because hidden channels are not included in the model.

The Authenticated Routing for Ad hoc Networks (ARAN) secure MANET routing protocol \cite{SanzgiriLDLSB05} is based on a public key signature scheme. In \cite{Godskesen06} a man-in-the-middle spoofing attack on ARAN is revealed by means of ProVerif \cite{Blanchet13}, an automatic cryptographic protocol verifier that transforms a protocol specification into Horn clauses. It is also explained how ARAN can be adapted to avoid this attack that constructs false routes. An analysis of this improved version of ARAN with AVISPA in \cite{PuraPB09} does not reveal any flaws. It is however reported that AVISPA does not terminate on some requirements in case of a network of five nodes. A later AVISPA analysis of the same protocol in \cite{BenettiMV10} does reveal attacks, in which routes are prevented from being discovered or existing routes are tampered with. In \cite{PuraBuchs15} an analysis of ARAN with the symbolic model checker AlPiNa \cite{HostettlerMLRB11} rediscovers all known attacks on this protocol.

In \cite{SosnovichGN13} the Open Shortest Path First (OSPF) routing protocol is analyzed using the model checker CBMC \cite{ClarkeKL04} for C programs, on networks of a few nodes. A spoofing attack is uncovered, as well as a attack in which the buffer of a node is filled with fake messages. Moreover, a known attack from \cite{NakiblyKGB12} is revealed. The same approach leads to the discovery of another attack in \cite{NakiblySMWE14}, in which nonexisting links are included in routing tables, giving rise to patches by most OSPF route vendors. In \cite{DarvilleHIP22} detailed {\sc Uppaal} models of OSPF are analyzed, leading to the automated detection of aforemenioned attacks.

Dedicated methods have been developed to facilitate the verification of security properties for MANET protocols by means of model checking. In \cite{AndelBY11} automation of model checking routing protocols is increased by evaluating all potential message sequences an attacker may use to change routing information during route discovery (in a given network). Additionally, automatic topology generation and reduction techniques are developed. Furthermore, a topology equivalence is defined to cluster network topologies with the same security vulnerabilities, allowing an exhaustive analysis for networks of six nodes. SPIN analyses of Ariadne and a secure version of DSR, using the Secure Routing Protocol from \cite{PapadimitratosHaas02}, show how an attack revealed in \cite{AcsBV06} can be discovered in an automated way. In \cite{CortierDD12} it is shown that for a large class of security properties of routing protocols, including validity of routes, it suffices to consider only five topologies consisting of four nodes. As examples, the secure version of DSR and the Secure, Disjoint, Multipath Source Routing (SDMSR) protocol \cite{BertonYLM06} are analyzed using ProVerif, again showing that an attack revealed in \cite{AcsBV06} can be discovered in an automated way. In the aforementioned paper \cite{SosnovichGN13} it is shown how a so-called abstract topology can, from a security perspective, capture a family of topologies of greater complexity.

\section{Modeling Frameworks for MANET Protocols}
\label{sec:frameworks}

Existing formal modeling languages have been extended and new formalisms have been developed to specifically target MANET networks. They played an important role in pushing forward the boundaries of formal analyses of MANET protocols. Typically, in these frameworks a message sent by a node is received only by the nodes in its range, and mobility is incorporated by letting neighborhoods of nodes evolve over time.

The majority of the frameworks targeting MANETs are in the form of a process calculus, in which the individual network nodes are specified in the form of algebraic terms and their behavior is defined by means of a formal semantics, which is often operational, meaning that transitions in the state space are defined on the basis of logical inference rules. A wide range of equivalences have been defined to distinguish such process behaviors.

This section starts with a general overview of process calculi for MANETs. Next, it is explained in some detail how broadcast communication has been modeled in these calculi. The section is completed with an overview of adaptations of other formalisms to specify MANET protocols, notably Object-Z, Abstract State Machines, and Petri Nets.

\subsection{Process Calculi}
\label{sec:pi-calculi}

In the Calculus of Broadcasting Systems with mobility (CBS$^{\#}$) \cite{NanzHankin06}, which targets MANET protocols, connectivity changes are captured in the semantics: every send event expresses which nodes are connected at that moment, and the transitions in the generated state space must satisfy a topology invariant. Process behavior is considered modulo what is called mediated equivalence, which identifies nodes with respect to their capabilities to store items from other nodes. A static analysis is proposed that by overapproximating network behavior allows to automatically analyze in how far an attacker can influence the network topology. A simplified version of SAODV, a secure variant of AODV, is specified in CBS$^{\#}$, and it is shown that this specification is not what is called topology consistent: even in the presence of a filter that rejects incorrect paths, a network under attack is not mediated equivalent to this network without an attacker. In \cite{ArnaudCD14} a process calculus inspired by CBS$^{\#}$ is defined to analyze the security of MANET routing protocols. Moreover, it is shown how to automatically search for network topologies that allow an attack, using constraint solving techniques. The approach is demonstrated on the aforementioned secure version of DSR.

In the Calculus of Mobile Ad hoc Networks (CMAN) \cite{Godskesen07,Godskesen08} as well as in the Calculus of Mobile ad hoc Networks (CMN) \cite{Merro09}, each network node is equipped with a location and communicates with other nodes using a spatially oriented broadcast. In CMAN arbitrary node mobility is expressed both in the semantics and syntactically through a static binding operator, while in CMN it is captured only in the semantics. For both process calculi, next to an operational semantics also a reduction semantics is provided, and the two semantics are shown to coincide with respect to a notion of weak bisimulation equivalence. The strength of CMAN is in \cite{Godskesen07} exemplified by means of the aforementioned attack on the ARAN secure routing protocol revealed in \cite{Godskesen06}. In \cite{MerroSibilio13} a variant of CMN is proposed that allows one to express different levels of trust.

In Restricted Broadcast Process Theory (RBPT) \cite{GhassemiFM08} connectivity changes are captured in the semantics, making sure that received messages in a broadcast are in sync with network connectivity. The notion of network bisimulation equivalence is introduced, which takes into account network connectivity. RBPT is provided with an equational theory in \cite{GhassemiFM10}. In \cite{GhassemiFM11} it is exemplified on a simplified version of the AODV protocol that this framework can be used to reason about MANET protocols on networks consisting of an unbounded number of nodes. A variant of RBPT in which communication is considered reliable, presented in \cite{GhassemiFokkink19}, aims at the detection of conceptual flaws in protocol designs that are not due to lossy communication. In the Calculus for Systems with Dynamic Topology (CSDT) \cite{KouzapasPhilippou11}, which is similar in spirit to RBPT, nodes can broadcast at two different transmission ranges. The belief of nodes about who are their neighbors is continuously updated. A hiding construct allows one to observe networks at different levels of abstraction. A theory of confluence is developed and the framework is applied to verify the Vasudevan-Kurose-Towsley leader election algorithm for MANETs \cite{VasudevanKT04}. In a stochastic extension of RBPT \cite{GhassemiTMF11}, delay functions are assigned to events, while the semantics captures the interplay of a MANET protocol with stochastic dynamic behavior of the data-link and physical networks layers. A continuous-time Markov chain, expressing the probability with which a certain state will change to another state, is derived using a novel notion of weak Markovian network bisimulation equivalence. This framework is imployed to analyze expected election times for the Vasudevan-Kurose-Towsley leader election algorithm.

A large body of work has been developed for the Algebra for Wireless Networks (AWN) \cite{FehnkerGHMPT12a}, which offers reliable local broadcast, a conditional unicast operator whose behavior depends on whether a message can be delivered, and rich datastructures. In \cite{FehnkerGHMPT12a,FehnkerGHMPT12b,HofnerGTPMF12} the core of AODV is modeled in AWN, including data handling aspects such as maintaining route tables. Moreover, properties such as loop freedom and packet delivery are (dis)proved, both manually and with {\sc Uppaal}, culminating in a detailed specification and proofs in \cite{GlabbeekHPT16}. This work is underpinned in \cite{BourkeGH14} by a formalization of proofs using the theorem prover Isabelle/HOL \cite{Paulson94}. A mechanization of the calculus AWN in Isabelle/HOL is presented in \cite{BourkeGH16}, together with a novel technique to lift global invariants from individual nodes to networks. By an automatic translation in \cite{GlabbeekHW18} from AWN to mCRL2, this formal verification toolset can be applied to AWN specifications.

A timed extension of AWN (T-AWN) is presented in \cite{BresGH16} and used to show that premature deletion of invalid routes and a too quick restart of a node after a reboot can lead to loops in AODV\@. Boundary conditions are given that resolve this problem. T-AWN is used in \cite{BarryGH20} to give an unambiguous specification of the OLSRv2 protocol. In \cite{DruryHW20} a detailed T-AWN specification is given of OSPF, providing an unambiguous interpretation of this standard. Moreover, the specification is translated to an {\sc Uppaal} model, serving as the basis for the aforementioned security analysis of OSPF in \cite{DarvilleHIP22}.

The Calculus of Wireless Systems (CWS) \cite{LaneseSangiorgi10} aims at modeling nonmobile wireless networks. An extension of CWS with mobility and time, called Timed Calculus for Mobile ad hoc Networks (TCMN), is proposed in \cite{WangLu12} and used to analyze two collision-avoidance protocols, Carrier Sense Multiple Access (CSMA) and Multiple Access with Collision Avoidance Receiver-Transmitter (MACA/R-T). It is shown that CSMA is not robust with regard to mobility while MACA/R-T is.

In \cite{ChenZhu23} the Calculus of the Internet of Thing \cite{LanotteMerro18} is extended with node mobility and broadcast communication. A transformation from the extended calculus to timed automata is presented, which is employed to verify six temporal properties for smart homes using {\sc Uppaal}.

\subsection{Extensions of the $\pi$-Calculus}
\label{sec:process-calculi}

The $\pi$-calculus \cite{MilnerPW92} is a process calculus in which channel names can be communicated along the channels themselves. This makes it a convenient platform for expressing mobility. A primary process calculus inspired by the $\pi$-calculus that takes the movement of nodes into account explicitly is the ambient calculus \cite{CardelliGordon00}. Locations are represented by a topology of boundaries, interaction between nodes is by shared location within a common boundary, and security is expressed by the inability to cross boundaries. Nomadic $\pi$ \cite{SewellWP98} adds communication primitives to the $\pi$-calculus for interaction between mobile agents in a two-level framework; the lower level defines location-dependent primitives, the higher layer defines location-independent primitives using underlying primitives. Nomadic $\pi$ serves as the basis for Nomadic Pict \cite{SewellWU10}, a distributed programming language for prototyping overlay network algorithms.

In three other extensions of the $\pi$-calculus, the $b\pi$-calculus \cite{EneMuntean01}, the Mobile Broadcasting System (MBS) calculus \cite{Prasad06}, and the $\omega$-calculus \cite{SinghRS10}, two nodes can communicate if they belong to the same group. Mobility is captured through the creation of new groups and allowing a node to move to another group. In \cite{EneMuntean01} a theoretical framework for this group communication paradigm is developed, including three behavioral equivalences. In \cite{SinghRS10} it is shown how one can verify a model against mobility scenarios using invariants that constrain mobility, which is applied to analyze AODV and the Vasudevan-Kurose-Towsley leader election algorithm. The psi-calculus \cite{BengtsonJPV11}, a parametric framework for extensions of the $\pi$-calculus, is in \cite{BorgstromHJRVPP11} provided with dynamic broadcast communication. This framework is demonstrated by verifying a reachability property for LUNAR.

Some extensions of the $\pi$-calculus target the security of MANETs. The Distributed Broadcast SPI-calculus (DBSPI) \cite{GodskesenHK09} combines elements of the Distributed $\pi$-calculus \cite{Hennessy07}, which adds a network layer and a primitive mobility construct to the $\pi$-calculus, and the Spi-calculus \cite{AbadiGordon97}, which is a cryptographic extension of the $\pi$-calculus. DBSPI offers a type system, i.e., a logical system of inference rules, that ensures correctness of authentication for MANET protocols. This framework is applied to prove the Mobile IP registration protocol \cite{Perkins96} correct. The bA$\pi$-calculus \cite{Godskesen10} adds mobility to the applied $\pi$-calculus \cite{AbadiFournet01}, an extension of the $\pi$-calculus that targets security protocols. The bA$\pi$-calculus allows one to reason about an unbounded number of nodes. Its reduction semantics, describing interactions between a sender and receivers, coincides with its operational semantics modulo a novel behavioral equivalence called barbed weak congruence. In \cite{ChretienDelaune13} a variant of the applied $\pi$-calculus is extended to analyze privacy-type properties for MANET routing protocols. Using this framework it is demonstrated that the ANonymous On-Demand Routing (ANODR) protocol \cite{KongHong03} violates source anonymity: an attacker can retrieve the identity of a node trying to establish a route.

\subsection{Modeling Broadcast Communication}

We now discuss in more detail, based on \cite[section 11.1]{FehnkerGHMPT13}, the ways in which broadcast communication has been modeled in the process calculi mentioned in the previous two sections. These calculi all include some form of broadcast communication in their operational semantics, in which a message emitted by a node can be received by multiple other nodes. Typically this is expressed by two symmetric operational rules such as\vspace{1.5mm}
$$\frac{M \trans{{\tt broadcast}(m)} M' \quad\quad N \trans{{\tt receive}(m)} N'}
  {M \| N \trans{{\tt broadcast}(m)} M' \| N'}
\qquad\qquad
  \frac{ M \trans{{\tt receive}(m)} M' \quad\quad N \trans{{\tt broadcast}(m)} N'}
  {M \| N \trans{{\tt broadcast}(m)} M' \| N'}$$\vspace{.5mm}
  
\noindent
which originate from the nonmobile Calculus of Broadcasting Systems (CBS) \cite{Prasad95}. Process terms $M$ and $N$ represent two subnetworks and $M\|N$ their composition. The premises at the top of the two operational rules express that in one of the subnetworks message $m$ is broadcast, while nodes in the other subnetwork can receive this message. The conclusion at the bottom of each rule expresses that then in the composed network these same nodes can receive $m$. In these rules the broadcast action in the conclusion is simply inherited from the broadcasting argument of the parallel composition, so that it remains available for the composition with yet another
receiver.

To capture the receipt of a broadcast messages $m$ properly, an additional operational rule is needed, expressing that multiple subnetworks can receive this message concurrently:\vspace{1.5mm}
$$\frac{M \trans{{\tt receive}(m)} M' \quad\quad N \trans{{\tt receive}(m)} N'}
  {M \| N \trans{{\tt receive}(m)} M' \| N'}$$\vspace{.5mm}
  
\noindent
This rule is missing in CMN and the $\omega$-calculus. In \cite{FehnkerGHMPT13} it is observed that this omission leads to technical problems in both calculi.

To model lossy communication in MANETS, the bA$\pi$-calculus, CMAN, CMN, the $\omega$-calculus, and RBPT contain operational rules to express that one subnetwork may nondeterministically miss a message $m$ broadcast by another subnetwork, even if they are within each other's ranges, such as the following two symmetric rules:\vspace{1.5mm}
$$\frac{M \trans{{\tt broadcast}(m)} M'}
  {M \| N \trans{{\tt broadcast}(m)} M' \| N}
\qquad\qquad\qquad
  \frac{N \trans{{\tt broadcast}(m)} N'}
  {M \| N \trans{{\tt broadcast}(m)} M \| N'}$$\vspace{.5mm}
  
\noindent 
A drawback of this approach is that it becomes impossible to verify properties such as ``if there is a path from some source node to some destination node and the topology remains stable, then data packets from the source will eventually reach their destination'' for routing protocols, because each packet may be lost on the way. AWN and RRBPT, a variant of RBPT, allow to express reliable communication and thus to verify such properties.

In the operational semantics of the b$\pi$-calculus, CBS$^{\#}$, CWS and an optional augmentation of AWN \cite{FehnkerGHMPT12a}, a broadcast message $m$ is only dropped by nodes that are not ready to receive it, which can be specified using a negative premise:\vspace{1.5mm}
$$\frac{M \trans{{\tt broadcast}(m)} M' \quad\quad N \ntrans{{\tt receive}(m)}}
  {M \| N \trans{{\tt broadcast}(m)} M' \| N}
\qquad\qquad
  \frac{ M \trans{{\tt receive}(m)} M' \quad\quad N \ntrans{{\tt receive}(m)}}
  {M \| N \trans{{\tt receive}(m)} M' \| N}$$\vspace{.5mm}
  
\noindent
and two more symmetric operational rules. The negative premise $N \ntrans{{\tt receive}(m)}$ expresses that process term $N$ is not able to perform the event ${\tt receive}(m)$. This makes these calculi nonblocking, meaning that a sender cannot be delayed in transmitting a message, even if potential recipients are not ready to receive it. The default version of AWN does allow for blocking, although it has facilities to prevent this in applications.

\subsection{Other Formalisms}

An early work where an existing formalism, not tailored to MANETS, is adapted to cope with mobility is \cite{RomanMP97}, which extends the UNITY proof logic for distributed systems \cite{ChandyMisra88}. Transient interactions are added to capture that nodes can communicate with each other only when they are within range, and the proof logic is extended to cover mobility. A baggage delivery system serves as running example.

Object-Z \cite{Smith00}, a state-oriented formal specification language with object-oriented concepts, is in \cite{Smith04} extended to dynamic networks whereby connections are mutable references between nodes. In \cite{WuSZ13} Object-Z is used to formalize the route discovery process in AODV, where broadcast communication is modeled by simultaneous changes in local variables of the sender and all of its connected receivers. Loop freedom is proved under a dynamic network topology. In \cite{KamaliPetre16} the {\sc Uppaal} model of the OLSR protocol from \cite{KamaliHKP15} is transposed to Event-B \cite{Abrial10}, to formally model this protocol through incremental stepwise refinements and prove functional properties with Event-B's interactive theorem prover Rodin. In \cite{KamaliPetre17} the experiences with this translation and the two models are the basis for some guidelines on when to use {\sc Uppaal} or Event-B for formal modeling and analysis. In \cite{FakhfakhTKM16} Event-B is combined with the evolving graphs formalism from \cite{Ferreira04} that aims to model dynamic graphs. The obtained framework to specify correct-by-construction MANET protocols is exemplified on a simple centralized counting algorithm.

In an Abstract State Machine (ASM) \cite{Gurevich93} states are data structures, equipped with functions and relations. A distributed ASM (DASM) \cite{BenczurGL03} is provided with a location service and position-based routing, to enable modeling MANET routing protocols. In \cite{GlasserGu05} this formalism is extended further with a logical topology and a protocol to let each node maintain the location information of its logical neighbors. In \cite{GlasserPrinz05} it is shown how the resulting formalism can be carried over to the specification language SDL \cite{RockstromSaracco82}. In \cite{BianchiPV14} AODV is modeled as a DASM and two correctness properties, starvation-freedom and that the correct packet is received back by the initiator of a route, are briefly argued based on this model. This work is in \cite{BianchiPV17} extended by proposing and formally proving correct a variant of AODV that prevents blackhole attacks, in which false information is sent toward the initiator.

An Algebraic Higher-Order (AHO) net \cite{HoffmannMossakowski02} allows the tokens in a Petri net to be dynamic structures, such as graphs or Petri nets. An architecture for AHO nets from \cite{BottoniDHM06} enables nodes in a MANET to collaborate effectively through workflows, by means of a coordination layer that maintains network connections and modifies the workflow schema at run-time. In \cite{PadbergHEMBE07} it is proposed to divide this architecture into three layers: a workflow layer, a mobility layer, and a team layer, which maps the activities of individual nodes to the workflow and mobility layers. In \cite{BiermannHP08} this approach is extended by making the mapping dynamic.

In \cite{ArchibaldKS21} it is investigated how topology control of wireless sensor networks can be modeled in two graph-based formalisms, Graph Transformation Systems \cite{EhrigPS73} and Bigraphical Reactive Systems \cite{Milner01}. In \cite{AlbalweAS24} bigraphs are used to model part of the Routing Protocol for low-power and Lossy Networks (RPL) \cite{AlexanderBVHPTLSKW12}, meant for wireless sensor networks. Bigraphs allow developers to draw the protocol updates. A model checking analysis of their (nonprobabilistic) model is performed using PRISM to verify a reachability property, discovery of optimal routes, and loop freedom.

\section{Mobility Models and Performance}
\label{sec:mobility}

A mobility model captures how nodes move through a MANET. This may be random or guided by some probability distribution. In \cite{CampBD02} it is shown that the choice of mobility model has a significant effect on the outcome of simulation experiments on MANETs to measure the performance of protocols. Different probabilistic formal frameworks have been proposed in which mobility models can be expressed explicitly.

An early work on mobility models is \cite{Patsouris01}, in which a pathset algebra is put forward, where a pathset is a graph structure of all paths from some source to some destination. It is coupled with an algebraic approach from \cite{RomanowskaSmith83} to model dynamic multi-agent systems. The framework is demonstrated on a small payroll application. In \cite{HofnerMcIver14} an algebra of routing tables for MANET routing protocols is developed that allows to algebraically reason about for instance packet delivery.

In \cite{GodskesenNanz09} a simple wireless process calculus is provided with mobility functions. Its operational semantics incorporates a notion of global time passing. This framework allows one to express and compare various mobility models with respect to weak simulation and bisimulation equivalences. In \cite{WuLZZ14} this mobility model is extended with group mobility and the ability to let movement patterns change over time, and this framework is applied to prove some properties for a simplified version of the Wireless HIerarchical Routing protocoL (WHIRL) \cite{PeiGHC99}, which includes group mobility. In \cite{SongGodskesen10} a mobile process calculus is proposed where messages may be lost with some probability. How the network distribution evolves depends on which nodes receive a broadcast message. Node mobility is ruled by a probabilistic function and the calculus is considered modulo a weak bisimulation equivalence that takes this mobility function into account. The framework is applied to analyze the probability of address collision for a simplifies version of \cite{CheshireAG05}, a protocol for assigning unique IP addresses to nodes in a self-configuring local area network. In \cite{SongGodskesen12} the mobility function is generalized by allowing it to change multiple connections at the same time. Furthermore, the process calculus is extended to express mobility models with stochastically timed behavior. This enhanced framework is demonstrated by a specification of the Vasudevan-Kurose-Towsley leader election algorithm.

\cite{GhassemiMF10} offers a framework to evaluate the performance of MANET protocols, capturing the interplay between stochastic behavior of protocols deployed at different network layers and the underlying topology. A link connectivity model specifies link up and down lifetimes. Transitions are annotated by network restrictions, expressing the topologies in which an event is possible. A continuous-time Markov chain is generated to evaluate performance by means of the probabilistic model checker PRISM \cite{HintonKNP06}.  A simple flooding protocol to find the largest ID in the network serves as running example. In \cite{MouradianAuge-Blum13} the unreliable nature of radio links is considered, where links may frequently appear and disappear. A method is presented to produce the set of topologies with a probability distribution and check the probability that a certain property holds. This method is applied to check some time properties on an {\sc Uppaal} model of f-MAC \cite{RoedigBS06}, a real-time protocol for access control of nodes in a wireless network to a shared medium. In \cite{KamaliKatoen20} the influence of uncertainty (such as packet loss rates, collisions) on the probability to establish short routes in AODV is analyzed using a formal model of probabilistic timed automata and the MODEST toolset \cite{BohnenkampDHK06}.

In \cite{FehnkerHKM13} a topology-based mobility model is proposed that abstracts away physical behavior and models mobility as probabilistic changes in the topology. These probabilities are distilled from either the random walk or the random waypoint model. The approach is applied, using {\sc Uppaal}, in a performance analysis of AODV and of the energy-efficient Lightweight Medium Access Control (LMAC) protocol \cite{HoeselHavinga04} for wireless sensor networks.

The Calculus for Wireless sensor networks from Quality perspective (CWQ) \cite{WuZhu15} is extended with mobility in \cite{WuZZ16} and the resulting calculus mCWQ is provided with an operational semantics. It offers a parametric framework to describe how node movement patterns evolve over time. A simple ad hoc network for vehicles serves as case study. In \cite{WuZX19} mCWQ is provided with a denotational semantics. In \cite{XieZWV19} a proof system for mCWQ is provided, based on Hoare Logic \cite{Hoare69} extended with some time primitives.

In \cite{LiuOM16}, different mobility models are specified in Real-Time Maude \cite{OlveczkyMeseguer00}, a specification language for real-time and hybrid systems based on rewriting logic. This framework is used to analyze the route discovery process in AODV and the Vasudevan-Kurose-Towsley leader election algorithm using Real-Time Maude's timed model checker \cite{LepriAO15}. This analysis uncovers a spurious behavior in the latter protocol, due to a subtle interplay between communication delays, node movement, and neighbor discovery.

PEBUM \cite{GallinaMR16} is an energy-aware process calculus for MANETs. A probabilistic bisimulation equivalence is defined to verify whether two networks exhibit the same observable behavior. Moreover, an energy-aware preorder is defined to compare the energy consumption of different but behaviorally equivalent networks. The framework is applied, using PRISM, to analyze how the performance of the LAR protocol depends on the network topology, and to compare the energy consumption of two error control protocols, stop-and-wait and go-back-N. This approach to turn a PEBUM specification into a discrete-time Markov chain and then apply PRISM is in \cite{GallinaMRHK13} applied to analyze energy consumption of different communication strategies for the gossip-based routing protocol from \cite{HaasHL06}. In \cite{BugliesiGHMR14} PEBUM is studied in the setting of a probabilistic reduction semantics from \cite{LaneseSangiorgi10} that takes into account interference caused by simultaneous transmissions of multiple nodes in overlapping transmission areas, leading to loss of messages at the receiver side. This offers a framework to analyze network connectivity and measure the level of interference in a MANET. As a case study it is proven that, under mild assumptions on node mobility, the LAR protocol has the same probability to discover a path as a simple flooding approach.

\section{Conclusions}
\label{sec:conclusions}

This survey aims to give an exhaustive overview of research performed in the context of rigorous formal methods for analyzing MANET protocols.
Formal modeling and analysis of MANETs has led to a significant body of research, including the development of innovative formalisms, semantics, and behavioral equivalence notions. These frameworks have been applied successfully to a wide range of MANET protocols, especially for routing, revealing many ambiguities and inconsistencies in their specifications as well as flaws in their designs, often leading to the adaptation of protocol standards. These imperfections would generally have been very difficult to spot with traditional methods such as visual inspection or testing an implementation of the standard, because they typically concern corner cases that rarely come to the surface and may require specific mobility scenarios.

A key aspect of these formalisms has been the invention of novel ways to express broadcast communication and mobility. Furthermore, the formal analysis of MANET protocols has benefited considerably from optimized verification techniques that specifically target MANETs. Symbolic methods have been of particular importance, because the exponential increase in possible mobility scenarios with the growth of a network seriously hampers model checking approaches, which require generation of the state space.

Summarizing, the following types of models and methods can be distinguished.
\begin{itemize}
\item
\emph{State-based model checking}: The system is represented as a collection of states and transitions between them, and model checking is employed to verify temporal logic formulas. General strengths of this approach are that it is relatively simple and intuitive, and well-supported by mature and automated tools such as SPIN. Typical use of this method in the context of this survey therefore is logical verification of protocol correctness on small MANET instances, including correctness of route discovery or packet delivery. General weaknesses are that only instances of networks can be verified, and the state explosion problem. The latter is particularly poignant for MANETs, due to poor scalability in case of many network nodes or dynamic topologies.
\item
\emph{Timed and hybrid}: Timed Automata, which include real-time clocks, can be used to model protocol timeouts and delays. Such models can be analyzed using a timed model checker such as {\sc Uppaal}. This is of importantance in the context of this survey because MANET routing often depends on timing constraints, e.g., route discovery completion within a time bound. Hybrid Automata include continuous variables which allow to combine discrete protocol logic with continuous mobility dynamics (e.g., node movement equations). The obvious weakness of both formalisms is that the state space explosion problem becomes even more severe. This can be alleviated by time abstraction or approximation of continuous behavior, but such measures tend to increase model complexity as well as the uncertainty of verification results.
\item
\emph{Probabilistic and stochastic}: Formalisms such as Probabilistic Automata and continuous-time Markov chains allow to represent uncertainty in transitions (e.g., mobility, link failure, packet loss), and thus capture randomness and reliability. This enables the employment of probabilistic model checkers such as PRISM for quantitative verification of robustness and reliability under uncertainty, e.g., whether a delivery probability is beyond a certain threshold. Typical use of such models in the context of MANET protocols is for analyzing reliability, packet loss, and expected route duration. General weaknesses are that this approach does not scale well to large state spaces and that it tends to require accurate probabilistic parameters.
\item
\emph{Petri net}: The system is modeled using events (transitions) and local states (places), with tokens modeling data moving through the network. Extensions such as Colored and Stochastic Petri nets provide more structure to tokens or probabilities to transitions. In the context of MANETs, tokens typically represent packets, while transitions model events like packet transmission or route discovery. General strengths are that this formalism allows to model parallelism, resource sharing, and synchronization of distributed and concurrent systems in a natural fashion. Moreover, the specification language is graphical and intuitive. Typical use of this formalism in the context of MANETS is for performance-oriented verification of concurrent behaviors in routing, and for studying packet flow, buffer management, and route establishment. General weaknesses are a limited expressiveness for complex logical properties, and that analysis can be computationally expensive.
\item
\emph{Process algebra}: The system is specified in an algebraic format, in which each network node is modeled as a concurrent process communicating through channels. General strengths of this framework are that it is good for reasoning about process interactions and synchronization. Strengths in the context of MANETs are that it naturally models concurrency and mobility, which is especially true for the $\pi$-calculus. The latter formalism is therefore particularly suited for MANETs where node connections change over time. Typical use of this method is to formally reason about communication and concurrency in MANET protocols, and to analyze protocol interaction patterns and message-passing correctness. General weaknesses are that the abstract mathematical formalism requires a steep learning curve, and that it can be hard to take into account quantitative aspects like timing or probabilities.
\item
\emph{Interactive proof assistant}: The user and the prover, such as Isabelle/HOL, collaborate to build a proof, whereby the user applies tactics or commands to break down goals into simpler subgoals and the prover checks each step for logical correctness and generates new subgoals if necessary. This typically entails the formulation and proof of invariants, e.g., sequence-number monotonicity for AODV. Existing libraries of lemmas and definitions can be exploited in the process. This approach avoids the state space explosion problem and can provide a formal correctness proof of MANET protocols for arbitrary networks, but often requires abstractions for mobility. Modeling realistic mobility and probabilistic loss in a proof assistant is possible but increases proof complexity significantly. The overall compute cost tends to be relatively low, but the human effort is high in terms of formalization and proof development, and requires skilled users. Building the right invariants is hard; here model checking can offer support through the discovery of counterexamples to refine invariants.
\end{itemize}

Process calculi sit between model checking and theorem proving in several respects: they give strong compositional and concurrency/mobility abstractions (very useful for MANETs), but they also incur non-trivial human cost (formal modeling + proof/analysis) and limited off-the-shelf automation compared with model checkers. They are especially attractive when you want protocol-level reasoning, compositional proofs (e.g., modular correctness, equivalence, refinement) and a natural treatment of dynamic link structure, but they are less convenient for brute-force counterexample search, tight timing/probabilistic analysis, or large concrete instance exploration. A process algebraic specification offers a useful first step toward a proof with an interactive proof assistant.

If priorities are modular design, correctness under dynamic topology, or reasoning about protocol refinements, invest in process calculi: they pay off conceptually and in proofs of equivalence/refinement. If priorities are fast bug discovery, timed/probabilistic metrics, or low-barrier automation, start with model checking. If you need machine-checked, parameterized guarantees for deployment or certification, plan for theorem proving — but bootstrap it from process calculus models and model checking counterexamples. For many MANET verification projects, a hybrid pipeline (model checking → process calculus → interactive proof assistant) may give the best cost/benefit balance.

The advancement of this research field is somewhat impeded by the fact that developed frameworks are often used only by their inventors. Preferably the rich body of work would lead to a single formalism that becomes the main vehicle for the formal analysis of MANET protocols. Process calculi seem to be best suited for this purpose, because they can conveniently express a combination of mobility as well as real-time, probabilistic, and security aspects. Furthermore, their algebraic nature allows to conveniently specify rich datastructures and efficiently perform both model checking and theorem proving analyses. Other important avenues for further research are techniques that target formal performance and security analyses of MANET protocols.

Formal analysis techniques regarding the functional correctness of MANET protocols have reached a level where complex, real-life protocol standards can be verified in a thorough fashion, exemplified by the detailed analysis of AODV in the process calculus AWN. A challenge is to develop the formalisms and techniques in such a way that they can be conveniently included and employed by protocol developers and software engineers in the development and implementation process of MANET protocol standards.


\bibliography{references}


\begin{thebibliography}{197}
\ifx \bisbn   \undefined \def \bisbn  #1{ISBN #1}\fi
\ifx \binits  \undefined \def \binits#1{#1}\fi
\ifx \bauthor  \undefined \def \bauthor#1{#1}\fi
\ifx \batitle  \undefined \def \batitle#1{#1}\fi
\ifx \bjtitle  \undefined \def \bjtitle#1{#1}\fi
\ifx \bvolume  \undefined \def \bvolume#1{\textbf{#1}}\fi
\ifx \byear  \undefined \def \byear#1{#1}\fi
\ifx \bissue  \undefined \def \bissue#1{#1}\fi
\ifx \bfpage  \undefined \def \bfpage#1{#1}\fi
\ifx \blpage  \undefined \def \blpage #1{#1}\fi
\ifx \burl  \undefined \def \burl#1{\textsf{#1}}\fi
\ifx \doiurl  \undefined \def \doiurl#1{\url{https://doi.org/#1}}\fi
\ifx \betal  \undefined \def \betal{\textit{et al.}}\fi
\ifx \binstitute  \undefined \def \binstitute#1{#1}\fi
\ifx \binstitutionaled  \undefined \def \binstitutionaled#1{#1}\fi
\ifx \bctitle  \undefined \def \bctitle#1{#1}\fi
\ifx \beditor  \undefined \def \beditor#1{#1}\fi
\ifx \bpublisher  \undefined \def \bpublisher#1{#1}\fi
\ifx \bbtitle  \undefined \def \bbtitle#1{#1}\fi
\ifx \bedition  \undefined \def \bedition#1{#1}\fi
\ifx \bseriesno  \undefined \def \bseriesno#1{#1}\fi
\ifx \blocation  \undefined \def \blocation#1{#1}\fi
\ifx \bsertitle  \undefined \def \bsertitle#1{#1}\fi
\ifx \bsnm \undefined \def \bsnm#1{#1}\fi
\ifx \bsuffix \undefined \def \bsuffix#1{#1}\fi
\ifx \bparticle \undefined \def \bparticle#1{#1}\fi
\ifx \barticle \undefined \def \barticle#1{#1}\fi
\bibcommenthead
\ifx \bconfdate \undefined \def \bconfdate #1{#1}\fi
\ifx \botherref \undefined \def \botherref #1{#1}\fi
\ifx \url \undefined \def \url#1{\textsf{#1}}\fi
\ifx \bchapter \undefined \def \bchapter#1{#1}\fi
\ifx \bbook \undefined \def \bbook#1{#1}\fi
\ifx \bcomment \undefined \def \bcomment#1{#1}\fi
\ifx \oauthor \undefined \def \oauthor#1{#1}\fi
\ifx \citeauthoryear \undefined \def \citeauthoryear#1{#1}\fi
\ifx \endbibitem  \undefined \def \endbibitem {}\fi
\ifx \bconflocation  \undefined \def \bconflocation#1{#1}\fi
\ifx \arxivurl  \undefined \def \arxivurl#1{\textsf{#1}}\fi
\csname PreBibitemsHook\endcsname

\bibitem[\protect\citeauthoryear{Jubin and Tornow}{1987}]{JubinTornow87}
\begin{barticle}
\bauthor{\bsnm{Jubin}, \binits{J.}},
\bauthor{\bsnm{Tornow}, \binits{J.D.}}:
\batitle{The {DARPA} packet radio network protocols}.
\bjtitle{Proc. IEEE}
\bvolume{75}(\bissue{1}),
\bfpage{21}--\blpage{32}
(\byear{1987})
\doiurl{10.1109/PROC.1987.13702}
\end{barticle}
\endbibitem

\bibitem[\protect\citeauthoryear{Shacham and
  Westcott}{1987}]{ShachamWestcott87}
\begin{barticle}
\bauthor{\bsnm{Shacham}, \binits{N.}},
\bauthor{\bsnm{Westcott}, \binits{J.}}:
\batitle{Future directions in packet radio architectures and protocols}.
\bjtitle{Proc. IEEE}
\bvolume{75}(\bissue{1}),
\bfpage{83}--\blpage{99}
(\byear{1987})
\doiurl{10.1109/PROC.1987.13707}
\end{barticle}
\endbibitem

\bibitem[\protect\citeauthoryear{Das et~al.}{2000}]{DasCY00}
\begin{barticle}
\bauthor{\bsnm{Das}, \binits{S.R.}},
\bauthor{\bsnm{Casta{\~{n}}eda}, \binits{R.}},
\bauthor{\bsnm{Yan}, \binits{J.}}:
\batitle{Simulation-based performance evaluation of routing protocols for
  mobile ad hoc networks}.
\bjtitle{Mob. Networks Appl.}
\bvolume{5}(\bissue{3}),
\bfpage{179}--\blpage{189}
(\byear{2000})
\doiurl{10.1023/A:1019108612308}
\end{barticle}
\endbibitem

\bibitem[\protect\citeauthoryear{Cavin et~al.}{2002}]{CavinSS02}
\begin{bchapter}
\bauthor{\bsnm{Cavin}, \binits{D.}},
\bauthor{\bsnm{Sasson}, \binits{Y.}},
\bauthor{\bsnm{Schiper}, \binits{A.}}:
\bctitle{On the accuracy of {MANET} simulators}.
In: \bbtitle{Proc. 1st Workshop on Principles of Mobile Computing (POMC)},
pp. \bfpage{38}--\blpage{43}.
\bpublisher{ACM},
\blocation{New York}
(\byear{2002}).
\doiurl{10.1145/584490.584499}
\end{bchapter}
\endbibitem

\bibitem[\protect\citeauthoryear{Roggenbach et~al.}{2022}]{RoggenbachSS22}
\begin{bchapter}
\bauthor{\bsnm{Roggenbach}, \binits{M.}},
\bauthor{\bsnm{Schlingloff}, \binits{B.-H.}},
\bauthor{\bsnm{Schneider}, \binits{G.}}:
\bctitle{Formal methods}.
In: \beditor{\bsnm{Roggenbach}, \binits{M.}},
\beditor{\bsnm{Cerone}, \binits{A.}},
\beditor{\bsnm{Schlingloff}, \binits{B.-H.}},
\beditor{\bsnm{Schneider}, \binits{G.}},
\beditor{\bsnm{Shaikh}, \binits{S.A.}} (eds.)
\bbtitle{Formal Methods for Software Engineering: Languages, Methods,
  Application Domains},
pp. \bfpage{1}--\blpage{46}.
\bpublisher{Springer},
\blocation{Heidelberg}
(\byear{2022}).
\doiurl{10.1007/978-3-030-38800-3}
\end{bchapter}
\endbibitem

\bibitem[\protect\citeauthoryear{Kulik et~al.}{2022}]{KulikDLMSTW22}
\begin{barticle}
\bauthor{\bsnm{Kulik}, \binits{T.}},
\bauthor{\bsnm{Dongol}, \binits{B.}},
\bauthor{\bsnm{Larsen}, \binits{P.G.}},
\bauthor{\bsnm{Macedo}, \binits{H.D.}},
\bauthor{\bsnm{Schneider}, \binits{S.}},
\bauthor{\bsnm{Tran{-}J{\o}rgensen}, \binits{P.W.V.}},
\bauthor{\bsnm{Woodcock}, \binits{J.}}:
\batitle{A survey of practical formal methods for security}.
\bjtitle{Formal Aspects Comput.}
\bvolume{34}(\bissue{1}),
\bfpage{1}--\blpage{39}
(\byear{2022})
\doiurl{10.1145/3522582}
\end{barticle}
\endbibitem

\bibitem[\protect\citeauthoryear{Clausen and Jacquet}{2003}]{ClausenJacquet03}
\begin{botherref}
\oauthor{\bsnm{Clausen}, \binits{T.H.}},
\oauthor{\bsnm{Jacquet}, \binits{P.}}:
Optimized Link State Routing Protocol (OLSR). RFC 3626.
IETF,
(2003).
IETF.
\url{http://www.ietf.org/rfc/rfc3626}
\end{botherref}
\endbibitem

\bibitem[\protect\citeauthoryear{Perkins and Royer}{1999}]{PerkinsRoyer99}
\begin{bchapter}
\bauthor{\bsnm{Perkins}, \binits{C.E.}},
\bauthor{\bsnm{Royer}, \binits{E.M.}}:
\bctitle{Ad-hoc on-demand distance vector routing}.
In: \bbtitle{Proc. 2nd Workshop on Mobile Computer Systems and Applications
  (WMCSA)},
pp. \bfpage{90}--\blpage{100}.
\bpublisher{IEEE},
\blocation{New York}
(\byear{1999}).
\doiurl{10.1109/MCSA.1999.749281}
\end{bchapter}
\endbibitem

\bibitem[\protect\citeauthoryear{Perkins et~al.}{2003}]{PerkinsRD03}
\begin{botherref}
\oauthor{\bsnm{Perkins}, \binits{C.E.}},
\oauthor{\bsnm{Royer}, \binits{E.M.}},
\oauthor{\bsnm{Das}, \binits{S.R.}}:
Ad Hoc On-Demand Distance Vector ({AODV}) Routing. RFC 3561.
IETF,
(2003).
IETF.
\url{https://datatracker.ietf.org/doc/html/rfc3561}
\end{botherref}
\endbibitem

\bibitem[\protect\citeauthoryear{Das and Dill}{2002}]{DasDill02}
\begin{bchapter}
\bauthor{\bsnm{Das}, \binits{S.}},
\bauthor{\bsnm{Dill}, \binits{D.L.}}:
\bctitle{Counter-example based predicate discovery in predicate abstraction}.
In: \bbtitle{Proc. 4th Conference on Formal Methods in Computer-Aided Design
  (FMCAD)}.
\bsertitle{LNCS},
vol. \bseriesno{2517},
pp. \bfpage{19}--\blpage{32}.
\bpublisher{Springer},
\blocation{Heidelberg}
(\byear{2002}).
\doiurl{10.1007/3-540-36126-X\_2}
\end{bchapter}
\endbibitem

\bibitem[\protect\citeauthoryear{Engler and
  Musuvathi}{2004}]{EnglerMusuvathi04}
\begin{bchapter}
\bauthor{\bsnm{Engler}, \binits{D.R.}},
\bauthor{\bsnm{Musuvathi}, \binits{M.}}:
\bctitle{Static analysis versus software model checking for bug finding}.
In: \bbtitle{Proc. 5th Conference on Verification, Model Checking, and Abstract
  Interpretation (VMCAI)}.
\bsertitle{LNCS},
vol. \bseriesno{2937},
pp. \bfpage{191}--\blpage{210}.
\bpublisher{Springer},
\blocation{Heidelberg}
(\byear{2004}).
\doiurl{10.1007/978-3-540-24622-0\_17}
\end{bchapter}
\endbibitem

\bibitem[\protect\citeauthoryear{Bhargavan et~al.}{2002}]{BhargavanOG02}
\begin{barticle}
\bauthor{\bsnm{Bhargavan}, \binits{K.}},
\bauthor{\bsnm{Obradovic}, \binits{D.}},
\bauthor{\bsnm{Gunter}, \binits{C.A.}}:
\batitle{Formal verification of standards for distance vector routing
  protocols}.
\bjtitle{JACM}
\bvolume{49}(\bissue{4}),
\bfpage{538}--\blpage{576}
(\byear{2002})
\doiurl{10.1145/581771.581775}
\end{barticle}
\endbibitem

\bibitem[\protect\citeauthoryear{Holzmann}{1997}]{Holzmann97}
\begin{barticle}
\bauthor{\bsnm{Holzmann}, \binits{G.J.}}:
\batitle{The model checker {SPIN}}.
\bjtitle{IEEE Trans. Software Eng.}
\bvolume{23}(\bissue{5}),
\bfpage{279}--\blpage{295}
(\byear{1997})
\doiurl{10.1109/32.588521}
\end{barticle}
\endbibitem

\bibitem[\protect\citeauthoryear{Gordon and Melham}{1993}]{GordonMelham93}
\begin{bbook}
\beditor{\bsnm{Gordon}, \binits{M.J.C.}},
\beditor{\bsnm{Melham}, \binits{T.F.}} (eds.):
\bbtitle{Introduction to HOL: A Theorem Proving Environment for Higher Order
  Logic}.
\bpublisher{Cambridge University Press},
\blocation{Cambridge}
(\byear{1993})
\end{bbook}
\endbibitem

\bibitem[\protect\citeauthoryear{Fehnker et~al.}{2013}]{FehnkerGHMPT13}
\begin{botherref}
\oauthor{\bsnm{Fehnker}, \binits{A.}},
\oauthor{\bsnm{{van Glabbeek}}, \binits{R.J.}},
\oauthor{\bsnm{H{\"{o}}fner}, \binits{P.}},
\oauthor{\bsnm{McIver}, \binits{A.}},
\oauthor{\bsnm{Portmann}, \binits{M.}},
\oauthor{\bsnm{Tan}, \binits{W.L.}}:
A process algebra for wireless mesh networks used for modelling, verifying and
  analysing {AODV}.
Technical Report 5513,
NICTA
(2013).
\doiurl{10.48550/arXiv.1312.7645}
\end{botherref}
\endbibitem

\bibitem[\protect\citeauthoryear{Wu et~al.}{2013}]{WuXZ13}
\begin{bchapter}
\bauthor{\bsnm{Wu}, \binits{X.}},
\bauthor{\bsnm{Xu}, \binits{Q.}},
\bauthor{\bsnm{Zhu}, \binits{H.}}:
\bctitle{Formal analysis of {AODV} using rely-guarantee}.
In: \bbtitle{Proc. 7th Symposium on Theoretical Aspects of Software Engineering
  (TASE)},
pp. \bfpage{45}--\blpage{48}.
\bpublisher{IEEE},
\blocation{New York}
(\byear{2013}).
\doiurl{10.1109/TASE.2013.14}
\end{bchapter}
\endbibitem

\bibitem[\protect\citeauthoryear{Zhou et~al.}{2009}]{ZhouYZW09}
\begin{bchapter}
\bauthor{\bsnm{Zhou}, \binits{M.}},
\bauthor{\bsnm{Yang}, \binits{H.}},
\bauthor{\bsnm{Zhang}, \binits{X.}},
\bauthor{\bsnm{Wang}, \binits{J.}}:
\bctitle{The proof of {AODV} loop freedom}.
In: \bbtitle{Proc. 1st Conference on Wireless Communications and Signal
  Processing (WCSP)},
pp. \bfpage{1}--\blpage{5}.
\bpublisher{ACM},
\blocation{New York}
(\byear{2009}).
\doiurl{10.1109/WCSP.2009.5371479}
\end{bchapter}
\endbibitem

\bibitem[\protect\citeauthoryear{{van Glabbeek} et~al.}{2013}]{GlabbeekHTP13}
\begin{bchapter}
\bauthor{\bsnm{{van Glabbeek}}, \binits{R.J.}},
\bauthor{\bsnm{H{\"{o}}fner}, \binits{P.}},
\bauthor{\bsnm{Tan}, \binits{W.L.}},
\bauthor{\bsnm{Portmann}, \binits{M.}}:
\bctitle{Sequence numbers do not guarantee loop freedom: {AODV} can yield
  routing loops}.
In: \bbtitle{Proc. 16th Conference on Modeling, Analysis and Simulation of
  Wireless and Mobile Systems (MSWiM)},
pp. \bfpage{91}--\blpage{100}.
\bpublisher{ACM},
\blocation{New York}
(\byear{2013}).
\doiurl{10.1145/2507924.2507943}
\end{bchapter}
\endbibitem

\bibitem[\protect\citeauthoryear{Jensen}{1981}]{Jensen81}
\begin{barticle}
\bauthor{\bsnm{Jensen}, \binits{K.}}:
\batitle{Coloured {Petri} nets and the invariant-method}.
\bjtitle{Theor. Comput. Sci.}
\bvolume{14},
\bfpage{317}--\blpage{336}
(\byear{1981})
\doiurl{10.1016/0304-3975(81)90049-9}
\end{barticle}
\endbibitem

\bibitem[\protect\citeauthoryear{Perkins et~al.}{2013}]{PerkinsRD02}
\begin{botherref}
\oauthor{\bsnm{Perkins}, \binits{C.E.}},
\oauthor{\bsnm{Ratliff}, \binits{S.}},
\oauthor{\bsnm{Dowdell}, \binits{J.}}:
Dynamic {MANET} On-Demand ({AODVv2}) Routing. Draft.
IETF,
(2013).
IETF.
\url{https://datatracker.ietf.org/doc/html/draft-ietf-manet-aodvv2-02}
\end{botherref}
\endbibitem

\bibitem[\protect\citeauthoryear{Xiong et~al.}{2002}]{XiongMT02}
\begin{bchapter}
\bauthor{\bsnm{Xiong}, \binits{C.}},
\bauthor{\bsnm{Murata}, \binits{T.}},
\bauthor{\bsnm{Tsai}, \binits{J.}}:
\bctitle{Modeling and simulation of routing protocol for mobile ad hoc networks
  using colored {Petri} nets}.
In: \bbtitle{Proc. Workshop on Formal Methods Applied to Defense Systems}.
\bsertitle{Conferences in Research and Practice in Information Technology},
vol. \bseriesno{12},
pp. \bfpage{145}--\blpage{153}.
\bpublisher{Australian Computer Society},
\blocation{Sydney}
(\byear{2002}).
\burl{https://dl.acm.org/doi/abs/10.5555/846335.846350}
\end{bchapter}
\endbibitem

\bibitem[\protect\citeauthoryear{Espensen et~al.}{2008}]{EspensenKK08}
\begin{bchapter}
\bauthor{\bsnm{Espensen}, \binits{K.L.}},
\bauthor{\bsnm{Kjeldsen}, \binits{M.K.}},
\bauthor{\bsnm{Kristensen}, \binits{L.M.}}:
\bctitle{Modelling and initial validation of the {DYMO} routing protocol for
  mobile ad-hoc networks}.
In: \bbtitle{Proc. 29th Conference on Applications and Theory of Petri Nets
  (PETRI NETS)}.
\bsertitle{LNCS},
vol. \bseriesno{5062},
pp. \bfpage{152}--\blpage{170}.
\bpublisher{Springer},
\blocation{Heidelberg}
(\byear{2008}).
\doiurl{10.1007/978-3-540-68746-7\_13}
\end{bchapter}
\endbibitem

\bibitem[\protect\citeauthoryear{Billington and Yuan}{2009}]{BillingtonYuan09}
\begin{barticle}
\bauthor{\bsnm{Billington}, \binits{J.}},
\bauthor{\bsnm{Yuan}, \binits{C.}}:
\batitle{On modelling and analysing the dynamic {MANET} on-demand {(DYMO)}
  routing protocol}.
\bjtitle{Trans. Petri Nets Other Model. Concurr.}
\bvolume{3},
\bfpage{98}--\blpage{126}
(\byear{2009})
\doiurl{10.1007/978-3-642-04856-2\_5}
\end{barticle}
\endbibitem

\bibitem[\protect\citeauthoryear{Edenhofer and
  H{\"{o}}fner}{2012}]{EdenhoferHofner12}
\begin{bchapter}
\bauthor{\bsnm{Edenhofer}, \binits{S.}},
\bauthor{\bsnm{H{\"{o}}fner}, \binits{P.}}:
\bctitle{Towards a rigorous analysis of {AODVv2} {(DYMO)}}.
In: \bbtitle{Proc. 20th Conference on Network Protocols (ICNP)},
pp. \bfpage{1}--\blpage{6}.
\bpublisher{IEEE},
\blocation{New York}
(\byear{2012}).
\doiurl{10.1109/ICNP.2012.6459942}
\end{bchapter}
\endbibitem

\bibitem[\protect\citeauthoryear{Saksena et~al.}{2008}]{SaksenaWJ08}
\begin{bchapter}
\bauthor{\bsnm{Saksena}, \binits{M.}},
\bauthor{\bsnm{Wibling}, \binits{O.}},
\bauthor{\bsnm{Jonsson}, \binits{B.}}:
\bctitle{Graph grammar modeling and verification of ad hoc routing protocols}.
In: \bbtitle{Proc. 14th Conference on Tools and Algorithms for the Construction
  and Analysis of Systems (TACAS)}.
\bsertitle{LNCS},
vol. \bseriesno{4963},
pp. \bfpage{18}--\blpage{32}.
\bpublisher{Springer},
\blocation{Heidelberg}
(\byear{2008}).
\doiurl{10.1007/978-3-540-78800-3\_3}
\end{bchapter}
\endbibitem

\bibitem[\protect\citeauthoryear{Namjoshi and
  Trefler}{2015}]{NamjoshiTrefler15b}
\begin{bchapter}
\bauthor{\bsnm{Namjoshi}, \binits{K.S.}},
\bauthor{\bsnm{Trefler}, \binits{R.J.}}:
\bctitle{Loop freedom in {AODVv2}}.
In: \bbtitle{Proc. 35th IFIP WG 6.1 Conference on Formal Techniques for
  Distributed Objects, Components, and Systems (FORTE)}.
\bsertitle{LNCS},
vol. \bseriesno{9039},
pp. \bfpage{98}--\blpage{112}.
\bpublisher{Springer},
\blocation{Heidelberg}
(\byear{2015}).
\doiurl{10.1007/978-3-319-19195-9\_7}
\end{bchapter}
\endbibitem

\bibitem[\protect\citeauthoryear{Yousefi et~al.}{2017}]{YousefiGK17}
\begin{barticle}
\bauthor{\bsnm{Yousefi}, \binits{B.}},
\bauthor{\bsnm{Ghassemi}, \binits{F.}},
\bauthor{\bsnm{Khosravi}, \binits{R.}}:
\batitle{Modeling and efficient verification of wireless ad hoc networks}.
\bjtitle{Formal Aspects Comput.}
\bvolume{29}(\bissue{6}),
\bfpage{1051}--\blpage{1086}
(\byear{2017})
\doiurl{10.1007/s00165-017-0429-z}
\end{barticle}
\endbibitem

\bibitem[\protect\citeauthoryear{Zakiuddin et~al.}{2003}]{ZakiuddinGWG03}
\begin{bchapter}
\bauthor{\bsnm{Zakiuddin}, \binits{I.}},
\bauthor{\bsnm{Goldsmith}, \binits{M.}},
\bauthor{\bsnm{Whittaker}, \binits{P.}},
\bauthor{\bsnm{Gardiner}, \binits{P.H.B.}}:
\bctitle{A methodology for model-checking ad-hoc networks}.
In: \bbtitle{Proc. 10th Workshop on Model Checking Software (SPIN)}.
\bsertitle{LNCS},
vol. \bseriesno{2648},
pp. \bfpage{181}--\blpage{196}.
\bpublisher{Springer},
\blocation{Heidelberg}
(\byear{2003}).
\doiurl{10.1007/3-540-44829-2\_12}
\end{bchapter}
\endbibitem

\bibitem[\protect\citeauthoryear{Lawrence}{2004}]{Lawrence04}
\begin{bchapter}
\bauthor{\bsnm{Lawrence}, \binits{J.}}:
\bctitle{Practical application of {CSP} and {FDR} to software design}.
In: \bbtitle{Communicating Sequential Processes: The First 25 Years}.
\bsertitle{LNCS},
vol. \bseriesno{3525},
pp. \bfpage{151}--\blpage{174}.
\bpublisher{Springer},
\blocation{Heidelberg}
(\byear{2004}).
\doiurl{10.1007/11423348\_9}
\end{bchapter}
\endbibitem

\bibitem[\protect\citeauthoryear{Jiang et~al.}{1998}]{JiangLT98}
\begin{botherref}
\oauthor{\bsnm{Jiang}, \binits{M.}},
\oauthor{\bsnm{Li}, \binits{J.}},
\oauthor{\bsnm{Tay}, \binits{Y.C.}}:
Cluster Based Routing Protocol (CBRP) Functional Specification. Draft.
IETF,
(1998).
IETF.
\url{https://tools.ietf.org/id/draft-ietf-manet-cbrp-spec-00.txt}
\end{botherref}
\endbibitem

\bibitem[\protect\citeauthoryear{Wibling et~al.}{2004}]{WiblingPP04}
\begin{bchapter}
\bauthor{\bsnm{Wibling}, \binits{O.}},
\bauthor{\bsnm{Parrow}, \binits{J.}},
\bauthor{\bsnm{Pears}, \binits{A.N.}}:
\bctitle{Automatized verification of ad hoc routing protocols}.
In: \bbtitle{Proc. 24th IFIP WG 6.1 Conference on Formal Techniques for
  Networked and Distributed Systems (FORTE)}.
\bsertitle{LNCS},
vol. \bseriesno{3235},
pp. \bfpage{343}--\blpage{358}.
\bpublisher{Springer},
\blocation{Heidelberg}
(\byear{2004}).
\doiurl{10.1007/978-3-540-30232-2\_22}
\end{bchapter}
\endbibitem

\bibitem[\protect\citeauthoryear{Tschudin et~al.}{2004}]{TschudinGRW04}
\begin{bchapter}
\bauthor{\bsnm{Tschudin}, \binits{C.}},
\bauthor{\bsnm{Gold}, \binits{R.}},
\bauthor{\bsnm{Rensfelt}, \binits{O.}},
\bauthor{\bsnm{Wibling}, \binits{O.}}:
\bctitle{{LUNAR}: A lightweight underlay network ad-hoc routing protocol and
  implementation}.
In: \bbtitle{Proc. 5th Conference on Next Generation Teletraffic and
  Wired/Wireless Advanced Networking (NEW2AN)},
pp. \bfpage{38}--\blpage{43}
(\byear{2004}).
\burl{https://www.researchgate.net/publication/228870993_LUNAR_a_lightweight_underlay_network_ad-hoc_routing_protocol_and_implementation}
\end{bchapter}
\endbibitem

\bibitem[\protect\citeauthoryear{{de Renesse} and
  Aghvami}{2004}]{RenesseAghvami04}
\begin{bchapter}
\bauthor{\bsnm{{de Renesse}}, \binits{R.}},
\bauthor{\bsnm{Aghvami}, \binits{A.H.}}:
\bctitle{Formal verification of ad-hoc routing protocols using {SPIN} model
  checker}.
In: \bbtitle{Proc. 12th Mediterranean Electrotechnical Conference (IEEE
  MELECON)},
pp. \bfpage{1177}--\blpage{1182}.
\bpublisher{IEEE},
\blocation{New York}
(\byear{2004}).
\doiurl{10.1109/MELCON.2004.1348275}
\end{bchapter}
\endbibitem

\bibitem[\protect\citeauthoryear{Khengar and Aghvami}{2001}]{KhengarAghvami01}
\begin{bchapter}
\bauthor{\bsnm{Khengar}, \binits{P.}},
\bauthor{\bsnm{Aghvami}, \binits{A.H.}}:
\bctitle{{WARP} - {T}he wireless adaptive routing protocol}.
In: \bbtitle{Proc. 10th IST Mobile Summit},
pp. \bfpage{480}--\blpage{485}
(\byear{2001})
\end{bchapter}
\endbibitem

\bibitem[\protect\citeauthoryear{Oleshchuk}{2005}]{Oleshchuk05}
\begin{barticle}
\bauthor{\bsnm{Oleshchuk}, \binits{V.A.}}:
\batitle{Modeling, specification and verification of ad-hoc sensor networks
  using {SPIN}}.
\bjtitle{Comput. Stand. Interfaces}
\bvolume{28}(\bissue{2}),
\bfpage{159}--\blpage{165}
(\byear{2005})
\doiurl{10.1016/j.csi.2005.01.017}
\end{barticle}
\endbibitem

\bibitem[\protect\citeauthoryear{Steele and Andel}{2012}]{SteeleAndel12}
\begin{bchapter}
\bauthor{\bsnm{Steele}, \binits{M.F.}},
\bauthor{\bsnm{Andel}, \binits{T.R.}}:
\bctitle{Modeling the optimized link-state routing protocol for verification}.
In: \bbtitle{Proc. Spring Simulation Multiconference (SpringSim)},
pp. \bfpage{1}--\blpage{8}.
\bpublisher{SCS/ACM},
\blocation{New York}
(\byear{2012}).
\burl{https://dl.acm.org/doi/abs/10.5555/2346616.2346651}
\end{bchapter}
\endbibitem

\bibitem[\protect\citeauthoryear{C{\^{a}}mara et~al.}{2007}]{CamaraLF07}
\begin{bchapter}
\bauthor{\bsnm{C{\^{a}}mara}, \binits{D.}},
\bauthor{\bsnm{Ferreira~Loureiro}, \binits{A.A.}},
\bauthor{\bsnm{Filali}, \binits{F.}}:
\bctitle{Methodology for formal verification of routing protocols for ad hoc
  wireless networks}.
In: \bbtitle{Proc. 8th Global Communications Conference (GLOBECOM)},
pp. \bfpage{705}--\blpage{709}.
\bpublisher{IEEE},
\blocation{New York}
(\byear{2007}).
\doiurl{10.1109/GLOCOM.2007.137}
\end{bchapter}
\endbibitem

\bibitem[\protect\citeauthoryear{Ko and Vaidya}{2000}]{KoVaidya00}
\begin{barticle}
\bauthor{\bsnm{Ko}, \binits{Y.}},
\bauthor{\bsnm{Vaidya}, \binits{N.H.}}:
\batitle{Location-aided routing {(LAR)} in mobile ad hoc networks}.
\bjtitle{Wirel. Networks}
\bvolume{6}(\bissue{4}),
\bfpage{307}--\blpage{321}
(\byear{2000})
\doiurl{10.1023/A:1019106118419}
\end{barticle}
\endbibitem

\bibitem[\protect\citeauthoryear{Basagni et~al.}{1998}]{BasagniCSW98}
\begin{bchapter}
\bauthor{\bsnm{Basagni}, \binits{S.}},
\bauthor{\bsnm{Chlamtac}, \binits{I.}},
\bauthor{\bsnm{Syrotiuk}, \binits{V.R.}},
\bauthor{\bsnm{Woodward}, \binits{B.A.}}:
\bctitle{A distance routing effect algorithm for mobility {(DREAM)}}.
In: \bbtitle{Proc. 4th Conference on Mobile Computing and Networking
  (MOBICOM)},
pp. \bfpage{76}--\blpage{84}.
\bpublisher{ACM},
\blocation{New York}
(\byear{1998}).
\doiurl{10.1145/288235.288254}
\end{bchapter}
\endbibitem

\bibitem[\protect\citeauthoryear{Singh et~al.}{2009}]{SinghRS09}
\begin{bchapter}
\bauthor{\bsnm{Singh}, \binits{A.}},
\bauthor{\bsnm{Ramakrishnan}, \binits{C.R.}},
\bauthor{\bsnm{Smolka}, \binits{S.A.}}:
\bctitle{Query-based model checking of ad hoc network protocols}.
In: \bbtitle{Proc. 20th Conference on Concurrency Theory (CONCUR)}.
\bsertitle{LNCS},
vol. \bseriesno{5710},
pp. \bfpage{603}--\blpage{619}.
\bpublisher{Springer},
\blocation{Heidelberg}
(\byear{2009}).
\doiurl{10.1007/978-3-642-04081-8\_40}
\end{bchapter}
\endbibitem

\bibitem[\protect\citeauthoryear{Ghassemi and
  Fokkink}{2016}]{GhassemiFokkink16}
\begin{barticle}
\bauthor{\bsnm{Ghassemi}, \binits{F.}},
\bauthor{\bsnm{Fokkink}, \binits{W.J.}}:
\batitle{Model checking mobile ad hoc networks}.
\bjtitle{Formal Methods Syst. Des.}
\bvolume{49}(\bissue{3}),
\bfpage{159}--\blpage{189}
(\byear{2016})
\doiurl{10.1007/s10703-016-0254-7}
\end{barticle}
\endbibitem

\bibitem[\protect\citeauthoryear{Groote et~al.}{2011}]{GrooteKSWW11}
\begin{barticle}
\bauthor{\bsnm{Groote}, \binits{J.F.}},
\bauthor{\bsnm{Keiren}, \binits{J.}},
\bauthor{\bsnm{Stappers}, \binits{F.P.M.}},
\bauthor{\bsnm{Wesselink}, \binits{W.}},
\bauthor{\bsnm{Willemse}, \binits{T.A.C.}}:
\batitle{Experiences in developing the {mCRL2} toolset}.
\bjtitle{Softw. Pract. Exp.}
\bvolume{41}(\bissue{2}),
\bfpage{143}--\blpage{153}
(\byear{2011})
\doiurl{10.1002/spe.1021}
\end{barticle}
\endbibitem

\bibitem[\protect\citeauthoryear{Nanz et~al.}{2010}]{NanzNN10}
\begin{barticle}
\bauthor{\bsnm{Nanz}, \binits{S.}},
\bauthor{\bsnm{Nielson}, \binits{F.}},
\bauthor{\bsnm{Nielson}, \binits{H.R.}}:
\batitle{Static analysis of topology-dependent broadcast networks}.
\bjtitle{Inf. Comput.}
\bvolume{208}(\bissue{2}),
\bfpage{117}--\blpage{139}
(\byear{2010})
\doiurl{10.1016/j.ic.2009.10.003}
\end{barticle}
\endbibitem

\bibitem[\protect\citeauthoryear{Namjoshi and
  Trefler}{2015}]{NamjoshiTrefler15a}
\begin{bchapter}
\bauthor{\bsnm{Namjoshi}, \binits{K.S.}},
\bauthor{\bsnm{Trefler}, \binits{R.J.}}:
\bctitle{Analysis of dynamic process networks}.
In: \bbtitle{Proc. 21st Conference on Tools and Algorithms for the Construction
  and Analysis of Systems (TACAS)}.
\bsertitle{LNCS},
vol. \bseriesno{9035},
pp. \bfpage{164}--\blpage{178}.
\bpublisher{Springer},
\blocation{Heidelberg}
(\byear{2015}).
\doiurl{10.1007/978-3-662-46681-0\_11}
\end{bchapter}
\endbibitem

\bibitem[\protect\citeauthoryear{Kojima et~al.}{2016}]{KojimaNT16}
\begin{bchapter}
\bauthor{\bsnm{Kojima}, \binits{H.}},
\bauthor{\bsnm{Nagashima}, \binits{Y.}},
\bauthor{\bsnm{Tsuchiya}, \binits{T.}}:
\bctitle{Model checking techniques for state space reduction in {MANET}
  protocol verification}.
In: \bbtitle{Proc. IPDPS Workshop on Advances in Parallel and Distributed
  Computational Models},
pp. \bfpage{509}--\blpage{516}.
\bpublisher{IEEE},
\blocation{New York}
(\byear{2016}).
\doiurl{10.1109/IPDPSW.2016.122}
\end{bchapter}
\endbibitem

\bibitem[\protect\citeauthoryear{Maag et~al.}{2008}]{MaagGC08}
\begin{barticle}
\bauthor{\bsnm{Maag}, \binits{S.}},
\bauthor{\bsnm{Grepet}, \binits{C.}},
\bauthor{\bsnm{Cavalli}, \binits{A.R.}}:
\batitle{A formal validation methodology for {MANET} routing protocols based on
  nodes' self similarity}.
\bjtitle{Comput. Commun.}
\bvolume{31}(\bissue{4}),
\bfpage{827}--\blpage{841}
(\byear{2008})
\doiurl{10.1016/j.comcom.2007.10.031}
\end{barticle}
\endbibitem

\bibitem[\protect\citeauthoryear{Djouvas et~al.}{2006}]{DjouvasGL06}
\begin{bchapter}
\bauthor{\bsnm{Djouvas}, \binits{C.}},
\bauthor{\bsnm{Griffeth}, \binits{N.D.}},
\bauthor{\bsnm{Lynch}, \binits{N.A.}}:
\bctitle{Testing self-similar networks}.
In: \bbtitle{Proc. 2nd Workshop on Model Based Testing (MBT)}.
\bsertitle{ENTCS},
vol. \bseriesno{164},
pp. \bfpage{67}--\blpage{82}.
\bpublisher{Elsevier},
\blocation{Amsterdam}
(\byear{2006}).
\doiurl{10.1016/j.entcs.2006.09.007}
\end{bchapter}
\endbibitem

\bibitem[\protect\citeauthoryear{Hu et~al.}{2007}]{HuMJ07}
\begin{botherref}
\oauthor{\bsnm{Hu}, \binits{Y.-C.}},
\oauthor{\bsnm{Maltz}, \binits{D.A.}},
\oauthor{\bsnm{Johnson}, \binits{D.B.}}:
The Dynamic Source Routing Protocol ({DSR}) for Mobile Ad Hoc Networks for
  IPv4. RFC 4728.
IETF,
(2007).
IETF.
\url{https://datatracker.ietf.org/doc/rfc4728/}
\end{botherref}
\endbibitem

\bibitem[\protect\citeauthoryear{Johnson et~al.}{2001}]{JohnsonMB01}
\begin{bchapter}
\bauthor{\bsnm{Johnson}, \binits{D.B.}},
\bauthor{\bsnm{Maltz}, \binits{D.A.}},
\bauthor{\bsnm{Broch}, \binits{J.}}:
\bctitle{{DSR}: The dynamic source routing protocol for multi-hop wireless ad
  hoc networks}.
In: \bbtitle{Ad Hoc Networking},
pp. \bfpage{139}--\blpage{172}.
\bpublisher{Addison-Wesley},
\blocation{Boston}
(\byear{2001}).
\burl{https://www.microsoft.com/en-us/research/publication/dsr-dynamic-source-routing-protocol-multi-hop-wireless-ad-hoc-networks/}
\end{bchapter}
\endbibitem

\bibitem[\protect\citeauthoryear{Belina and Hogrefe}{1989}]{BelinaHogrefe89}
\begin{barticle}
\bauthor{\bsnm{Belina}, \binits{F.}},
\bauthor{\bsnm{Hogrefe}, \binits{D.}}:
\batitle{The {CCITT}-specification and description language {SDL}}.
\bjtitle{Comput. Netw. ISDN Syst.}
\bvolume{16}(\bissue{4}),
\bfpage{311}--\blpage{341}
(\byear{1989})
\doiurl{10.1016/0169-7552(89)90078-0}
\end{barticle}
\endbibitem

\bibitem[\protect\citeauthoryear{Andr{\'{e}}s et~al.}{2009}]{AndresMCMN09}
\begin{bchapter}
\bauthor{\bsnm{Andr{\'{e}}s}, \binits{C.}},
\bauthor{\bsnm{Maag}, \binits{S.}},
\bauthor{\bsnm{Cavalli}, \binits{A.R.}},
\bauthor{\bsnm{Merayo}, \binits{M.G.}},
\bauthor{\bsnm{N{\'{u}}{\~{n}}ez}, \binits{M.}}:
\bctitle{Analysis of the {OLSR} protocol by using formal passive testing}.
In: \bbtitle{Proc. 16th Asia-Pacific Software Engineering Conference (APSEC)},
pp. \bfpage{152}--\blpage{159}.
\bpublisher{IEEE},
\blocation{New York}
(\byear{2009}).
\doiurl{10.1109/APSEC.2009.37}
\end{bchapter}
\endbibitem

\bibitem[\protect\citeauthoryear{Lalanne and Maag}{2013}]{LalanneMaag13}
\begin{bchapter}
\bauthor{\bsnm{Lalanne}, \binits{F.}},
\bauthor{\bsnm{Maag}, \binits{S.}}:
\bctitle{Datamonitor - {A} formal approach for passively testing a {MANET}
  routing protocol}.
In: \bbtitle{Proc. 9th Wireless Communications and Mobile Computing Conference
  (IWCMC)},
pp. \bfpage{207}--\blpage{212}.
\bpublisher{IEEE},
\blocation{New York}
(\byear{2013}).
\doiurl{10.1109/IWCMC.2013.6583560}
\end{bchapter}
\endbibitem

\bibitem[\protect\citeauthoryear{Hedrick}{1988}]{Hedrick88}
\begin{botherref}
\oauthor{\bsnm{Hedrick}, \binits{C.}}:
Routing Information Protocol. RFC 1058.
IETF,
(1988).
IETF.
\url{https://datatracker.ietf.org/doc/html/rfc1058}
\end{botherref}
\endbibitem

\bibitem[\protect\citeauthoryear{Larsen et~al.}{1997}]{LarsenPY97}
\begin{barticle}
\bauthor{\bsnm{Larsen}, \binits{K.G.}},
\bauthor{\bsnm{Pettersson}, \binits{P.}},
\bauthor{\bsnm{Yi}, \binits{W.}}:
\batitle{{{\sc Uppaal}} in a nutshell}.
\bjtitle{Int. J. Softw. Tools Technol. Transf.}
\bvolume{1}(\bissue{1-2}),
\bfpage{134}--\blpage{152}
(\byear{1997})
\doiurl{10.1007/s100090050010}
\end{barticle}
\endbibitem

\bibitem[\protect\citeauthoryear{Wibling et~al.}{2005}]{WiblingPP05}
\begin{bchapter}
\bauthor{\bsnm{Wibling}, \binits{O.}},
\bauthor{\bsnm{Parrow}, \binits{J.}},
\bauthor{\bsnm{Pears}, \binits{A.N.}}:
\bctitle{Ad hoc routing protocol verification through broadcast abstraction}.
In: \bbtitle{Proc. 25th IFIP WG 6.1 Conference on Formal Techniques for
  Networked and Distributed Systems (FORTE)}.
\bsertitle{LNCS},
vol. \bseriesno{3731},
pp. \bfpage{128}--\blpage{142}.
\bpublisher{Springer},
\blocation{Heidelberg}
(\byear{2005}).
\doiurl{10.1007/11562436\_11}
\end{bchapter}
\endbibitem

\bibitem[\protect\citeauthoryear{Chiyangwa and
  Kwiatkowska}{2005}]{ChiyangwaKwiatkowska05}
\begin{bchapter}
\bauthor{\bsnm{Chiyangwa}, \binits{S.}},
\bauthor{\bsnm{Kwiatkowska}, \binits{M.Z.}}:
\bctitle{A timing analysis of {AODV}}.
In: \bbtitle{Proc. 7th IFIP WG 6.1 Conference on Formal Methods for Open
  Object-Based Distributed Systems (FMOODS)}.
\bsertitle{LNCS},
vol. \bseriesno{3535},
pp. \bfpage{306}--\blpage{321}.
\bpublisher{Springer},
\blocation{Heidelberg}
(\byear{2005}).
\doiurl{10.1007/11494881\_20}
\end{bchapter}
\endbibitem

\bibitem[\protect\citeauthoryear{Chaudhary et~al.}{2017}]{ChaudharyFM17}
\begin{bchapter}
\bauthor{\bsnm{Chaudhary}, \binits{K.}},
\bauthor{\bsnm{Fehnker}, \binits{A.}},
\bauthor{\bsnm{Mehta}, \binits{V.}}:
\bctitle{Modelling, verification, and comparative performance analysis of the
  {B.A.T.M.A.N.} protocol}.
In: \bbtitle{Proc. 2nd Workshop on Models for Formal Analysis of Real Systems
  (MARS@ETAPS)}.
\bsertitle{{EPTCS}},
vol. \bseriesno{244},
pp. \bfpage{53}--\blpage{65}
(\byear{2017}).
\doiurl{10.4204/EPTCS.244.3}
\end{bchapter}
\endbibitem

\bibitem[\protect\citeauthoryear{Furlan}{2012}]{Furlan12}
\begin{botherref}
\oauthor{\bsnm{Furlan}, \binits{D.}}:
Improving {B.A.T.M.A.N.} routing stability and performance.
Master's thesis,
Universtit\`a degli Studi di Trento
(2012).
\url{https://www.semanticscholar.org/paper/Improving-B.A.T.M.A.N.-Routing-Stability-and-Cigno-Furlan/e73b1615721f5c2adb4d0f2129d10419b133e7f7}
\end{botherref}
\endbibitem

\bibitem[\protect\citeauthoryear{Neumann et~al.}{2008}]{NeumannALW08}
\begin{botherref}
\oauthor{\bsnm{Neumann}, \binits{A.}},
\oauthor{\bsnm{Aichele}, \binits{C.}},
\oauthor{\bsnm{Lindner}, \binits{M.}},
\oauthor{\bsnm{Wunderlich}, \binits{S.}}:
Better Approach To Mobile Ad-hoc Networking (B.A.T.M.A.N.). Draft.
IETF,
(2008).
IETF.
\url{https://datatracker.ietf.org/doc/html/draft-wunderlich-openmesh-manet-routing-00}
\end{botherref}
\endbibitem

\bibitem[\protect\citeauthoryear{Fehnker et~al.}{2018}]{FehnkerCM18}
\begin{bchapter}
\bauthor{\bsnm{Fehnker}, \binits{A.}},
\bauthor{\bsnm{Chaudhary}, \binits{K.}},
\bauthor{\bsnm{Mehta}, \binits{V.}}:
\bctitle{An even better approach - {I}mproving the {B.A.T.M.A.N.} protocol
  through formal modelling and analysis}.
In: \bbtitle{Proc. 10th NASA Formal Methods Symposium (NFM)}.
\bsertitle{LNCS},
vol. \bseriesno{10811},
pp. \bfpage{164}--\blpage{178}.
\bpublisher{Springer},
\blocation{Heidelberg}
(\byear{2018}).
\doiurl{10.1007/978-3-319-77935-5\_12}
\end{bchapter}
\endbibitem

\bibitem[\protect\citeauthoryear{Kamali et~al.}{2015}]{KamaliHKP15}
\begin{bchapter}
\bauthor{\bsnm{Kamali}, \binits{M.}},
\bauthor{\bsnm{H{\"{o}}fner}, \binits{P.}},
\bauthor{\bsnm{Kamali}, \binits{M.}},
\bauthor{\bsnm{Petre}, \binits{L.}}:
\bctitle{Formal analysis of proactive, distributed routing}.
In: \bbtitle{Proc. 13th Conference on Software Engineering and Formal Methods
  (SEFM)}.
\bsertitle{LNCS},
vol. \bseriesno{9276},
pp. \bfpage{175}--\blpage{189}.
\bpublisher{Springer},
\blocation{Heidelberg}
(\byear{2015}).
\doiurl{10.1007/978-3-319-22969-0\_13}
\end{bchapter}
\endbibitem

\bibitem[\protect\citeauthoryear{Kamali and Petre}{2015}]{KamaliPetre15}
\begin{bchapter}
\bauthor{\bsnm{Kamali}, \binits{M.}},
\bauthor{\bsnm{Petre}, \binits{L.}}:
\bctitle{Improved recovery for proactive, distributed routing}.
In: \bbtitle{Proc. 20th Conference on Engineering of Complex Computer Systems
  (ICECCS)},
pp. \bfpage{178}--\blpage{181}.
\bpublisher{IEEE},
\blocation{New York}
(\byear{2015}).
\doiurl{10.1109/ICECCS.2015.27}
\end{bchapter}
\endbibitem

\bibitem[\protect\citeauthoryear{Mouradian and
  Aug{\'{e}}{-}Blum}{2014}]{MouradianAuge-Blum14}
\begin{bchapter}
\bauthor{\bsnm{Mouradian}, \binits{A.}},
\bauthor{\bsnm{Aug{\'{e}}{-}Blum}, \binits{I.}}:
\bctitle{Formal verification of real-time wireless sensor networks protocols:
  Scaling up}.
In: \bbtitle{Proc. 26th Euromicro Conference on Real-Time Systems (ECRTS)},
pp. \bfpage{41}--\blpage{50}.
\bpublisher{IEEE},
\blocation{New York}
(\byear{2014}).
\doiurl{10.1109/ECRTS.2014.12}
\end{bchapter}
\endbibitem

\bibitem[\protect\citeauthoryear{Schmitt and Roedig}{2005}]{SchmittRoedig05}
\begin{bchapter}
\bauthor{\bsnm{Schmitt}, \binits{J.B.}},
\bauthor{\bsnm{Roedig}, \binits{U.}}:
\bctitle{Sensor network calculus - {A} framework for worst case analysis}.
In: \bbtitle{Proc. 1st Conference in Distributed Computing in Sensor Systems
  (DCOSS)}.
\bsertitle{LNCS},
vol. \bseriesno{3560},
pp. \bfpage{141}--\blpage{154}.
\bpublisher{Springer},
\blocation{Heidelberg}
(\byear{2005}).
\doiurl{10.1007/11502593\_13}
\end{bchapter}
\endbibitem

\bibitem[\protect\citeauthoryear{Mouradian et~al.}{2014}]{MouradianAV14}
\begin{barticle}
\bauthor{\bsnm{Mouradian}, \binits{A.}},
\bauthor{\bsnm{Aug{\'{e}}{-}Blum}, \binits{I.}},
\bauthor{\bsnm{Valois}, \binits{F.}}:
\batitle{{RTXP}: A localized real-time {MAC}-routing protocol for wireless
  sensor networks}.
\bjtitle{Comput. Networks}
\bvolume{67},
\bfpage{43}--\blpage{59}
(\byear{2014})
\doiurl{10.1016/j.comnet.2014.03.020}
\end{barticle}
\endbibitem

\bibitem[\protect\citeauthoryear{Hammal et~al.}{2017}]{HammalSBA17}
\begin{bchapter}
\bauthor{\bsnm{Hammal}, \binits{Y.}},
\bauthor{\bsnm{Seddiki}, \binits{M.}},
\bauthor{\bsnm{Bencha{\"{\i}}ba}, \binits{M.}},
\bauthor{\bsnm{Abdelli}, \binits{A.}}:
\bctitle{Formal specification and analysis of a cross-layer overlay {P2P}
  construction protocol over {MANETs}}.
In: \bbtitle{Proc.\ 18th Wireless Communications and Networking Conference
  (WCNC)},
pp. \bfpage{1}--\blpage{6}.
\bpublisher{IEEE},
\blocation{New York}
(\byear{2017}).
\doiurl{10.1109/WCNC.2017.7925622}
\end{bchapter}
\endbibitem

\bibitem[\protect\citeauthoryear{H{\"{o}}fner and
  McIver}{2013}]{HofnerMcIver13}
\begin{bchapter}
\bauthor{\bsnm{H{\"{o}}fner}, \binits{P.}},
\bauthor{\bsnm{McIver}, \binits{A.}}:
\bctitle{Statistical model checking of wireless mesh routing protocols}.
In: \bbtitle{Proc. 5th NASA Formal Methods Symposium (NFM)}.
\bsertitle{LNCS},
vol. \bseriesno{7871},
pp. \bfpage{322}--\blpage{336}.
\bpublisher{Springer},
\blocation{Heidelberg}
(\byear{2013}).
\doiurl{10.1007/978-3-642-38088-4\_22}
\end{bchapter}
\endbibitem

\bibitem[\protect\citeauthoryear{David et~al.}{2015}]{DavidLLMP15}
\begin{barticle}
\bauthor{\bsnm{David}, \binits{A.}},
\bauthor{\bsnm{Larsen}, \binits{K.G.}},
\bauthor{\bsnm{Legay}, \binits{A.}},
\bauthor{\bsnm{Miku\u{c}ionis}, \binits{M.}},
\bauthor{\bsnm{Poulsen}, \binits{D.B.}}:
\batitle{{{\sc Uppaal}} {SMC} tutorial}.
\bjtitle{Int. J. Softw. Tools Technol. Transf.}
\bvolume{17}(\bissue{4}),
\bfpage{397}--\blpage{415}
(\byear{2015})
\doiurl{10.1007/s10009-014-0361-y}
\end{barticle}
\endbibitem

\bibitem[\protect\citeauthoryear{Dal~Corso et~al.}{2015}]{DalCorsoMM15}
\begin{bchapter}
\bauthor{\bsnm{Dal~Corso}, \binits{A.}},
\bauthor{\bsnm{Macedonio}, \binits{D.}},
\bauthor{\bsnm{Merro}, \binits{M.}}:
\bctitle{Statistical model checking of ad hoc routing protocols in lossy grid
  networks}.
In: \bbtitle{Proc. 7th NASA Formal Methods Symposium (NFM)}.
\bsertitle{LNCS},
vol. \bseriesno{9058},
pp. \bfpage{112}--\blpage{126}.
\bpublisher{Springer},
\blocation{Heidelberg}
(\byear{2015}).
\doiurl{10.1007/978-3-319-17524-9\_9}
\end{bchapter}
\endbibitem

\bibitem[\protect\citeauthoryear{Kamali et~al.}{2018}]{KamaliMD18}
\begin{bchapter}
\bauthor{\bsnm{Kamali}, \binits{M.}},
\bauthor{\bsnm{Merro}, \binits{M.}},
\bauthor{\bsnm{Dal~Corso}, \binits{A.}}:
\bctitle{{AODVv2}: Performance vs. loop freedom}.
In: \bbtitle{Proc. 44th Conference on Current Trends in Theory and Practice of
  Computer Science (SOFSEM)}.
\bsertitle{LNCS},
vol. \bseriesno{10706},
pp. \bfpage{337}--\blpage{350}.
\bpublisher{Springer},
\blocation{Heidelberg}
(\byear{2018}).
\doiurl{10.1007/978-3-319-73117-9\_24}
\end{bchapter}
\endbibitem

\bibitem[\protect\citeauthoryear{Kamali and Fehnker}{2018}]{KamaliFehnker18}
\begin{bchapter}
\bauthor{\bsnm{Kamali}, \binits{M.}},
\bauthor{\bsnm{Fehnker}, \binits{A.}}:
\bctitle{Adaptive formal framework for {WMN} routing protocols}.
In: \bbtitle{Proc. 15th Conference on Formal Aspects of Component Software
  (FACS)}.
\bsertitle{LNCS},
vol. \bseriesno{11222},
pp. \bfpage{175}--\blpage{195}.
\bpublisher{Springer},
\blocation{Heidelberg}
(\byear{2018}).
\doiurl{10.1007/978-3-030-02146-7\_9}
\end{bchapter}
\endbibitem

\bibitem[\protect\citeauthoryear{Herberg and Clausen}{2010}]{HerbergClausen10}
\begin{barticle}
\bauthor{\bsnm{Herberg}, \binits{U.}},
\bauthor{\bsnm{Clausen}, \binits{T.H.}}:
\batitle{Security issues in the optimized link state routing protocol version 2
  ({OLSRv2})}.
\bjtitle{Int. J. Netw. Secur. Appl.}
\bvolume{2}(\bissue{2}),
\bfpage{162}--\blpage{181}
(\byear{2010})
\doiurl{10.5121/ijnsa.2010.2213}
\end{barticle}
\endbibitem

\bibitem[\protect\citeauthoryear{Thayer et~al.}{1999}]{ThayerHG99}
\begin{barticle}
\bauthor{\bsnm{Thayer}, \binits{F.J.}},
\bauthor{\bsnm{Herzog}, \binits{J.C.}},
\bauthor{\bsnm{Guttman}, \binits{J.D.}}:
\batitle{Strand spaces: Proving security protocols correct}.
\bjtitle{J. Comput. Secur.}
\bvolume{7}(\bissue{1}),
\bfpage{191}--\blpage{230}
(\byear{1999})
\doiurl{10.3233/JCS-1999-72-304}
\end{barticle}
\endbibitem

\bibitem[\protect\citeauthoryear{Song}{1999}]{Song99}
\begin{bchapter}
\bauthor{\bsnm{Song}, \binits{D.X.}}:
\bctitle{Athena: {A} new efficient automatic checker for security protocol
  analysis}.
In: \bbtitle{Proc. 12th Computer Security Foundations Workshop (CSFW)},
pp. \bfpage{192}--\blpage{202}.
\bpublisher{IEEE},
\blocation{New York}
(\byear{1999}).
\doiurl{10.1109/CSFW.1999.779773}
\end{bchapter}
\endbibitem

\bibitem[\protect\citeauthoryear{Yang and Baras}{2003}]{YangBaras03}
\begin{bchapter}
\bauthor{\bsnm{Yang}, \binits{S.}},
\bauthor{\bsnm{Baras}, \binits{J.S.}}:
\bctitle{Modeling vulnerabilities of ad hoc routing protocols}.
In: \bbtitle{Proc. 1st Workshop on Security of Ad Hoc and Sensor Networks
  (SASN)},
pp. \bfpage{12}--\blpage{20}.
\bpublisher{ACM},
\blocation{New York}
(\byear{2003}).
\doiurl{10.1145/986858.986861}
\end{bchapter}
\endbibitem

\bibitem[\protect\citeauthoryear{Hu et~al.}{2005}]{HuPJ05}
\begin{barticle}
\bauthor{\bsnm{Hu}, \binits{Y.}},
\bauthor{\bsnm{Perrig}, \binits{A.}},
\bauthor{\bsnm{Johnson}, \binits{D.B.}}:
\batitle{Ariadne: {A} secure on-demand routing protocol for ad hoc networks}.
\bjtitle{Wirel. Networks}
\bvolume{11}(\bissue{1-2}),
\bfpage{21}--\blpage{38}
(\byear{2005})
\doiurl{10.1007/s11276-004-4744-y}
\end{barticle}
\endbibitem

\bibitem[\protect\citeauthoryear{{\'{A}}cs et~al.}{2006}]{AcsBV06}
\begin{barticle}
\bauthor{\bsnm{{\'{A}}cs}, \binits{G.}},
\bauthor{\bsnm{Butty{\'{a}}n}, \binits{L.}},
\bauthor{\bsnm{Vajda}, \binits{I.}}:
\batitle{Provably secure on-demand source routing in mobile ad hoc networks}.
\bjtitle{IEEE Trans. Mob. Comput.}
\bvolume{5}(\bissue{11}),
\bfpage{1533}--\blpage{1546}
(\byear{2006})
\doiurl{10.1109/TMC.2006.170}
\end{barticle}
\endbibitem

\bibitem[\protect\citeauthoryear{Burmester and {de
  Medeiros}}{2009}]{BurmesterM09}
\begin{barticle}
\bauthor{\bsnm{Burmester}, \binits{M.}},
\bauthor{\bsnm{{de Medeiros}}, \binits{B.}}:
\batitle{On the security of route discovery in {MANETs}}.
\bjtitle{IEEE Trans. Mob. Comput.}
\bvolume{8}(\bissue{9}),
\bfpage{1180}--\blpage{1188}
(\byear{2009})
\doiurl{10.1109/TMC.2009.13}
\end{barticle}
\endbibitem

\bibitem[\protect\citeauthoryear{Benetti et~al.}{2010}]{BenettiMV10}
\begin{bchapter}
\bauthor{\bsnm{Benetti}, \binits{D.}},
\bauthor{\bsnm{Merro}, \binits{M.}},
\bauthor{\bsnm{Vigan{\`{o}}}, \binits{L.}}:
\bctitle{Model checking ad hoc network routing protocols: {ARAN} vs.
  {endairA}}.
In: \bbtitle{Proc. 8th Conference on Software Engineering and Formal Methods
  (SEFM)},
pp. \bfpage{191}--\blpage{202}.
\bpublisher{IEEE},
\blocation{New York}
(\byear{2010}).
\doiurl{10.1109/SEFM.2010.24}
\end{bchapter}
\endbibitem

\bibitem[\protect\citeauthoryear{Armando
  et~al.}{2005}]{ArmandoBBCCCDHKMMORSTVV05}
\begin{bchapter}
\bauthor{\bsnm{Armando}, \binits{A.}},
\bauthor{\bsnm{Basin}, \binits{D.A.}},
\bauthor{\bsnm{Boichut}, \binits{Y.}},
\bauthor{\bsnm{Chevalier}, \binits{Y.}},
\bauthor{\bsnm{Compagna}, \binits{L.}},
\bauthor{\bsnm{Cu{\'{e}}llar}, \binits{J.}},
\bauthor{\bsnm{Hankes~Drielsma}, \binits{P.}},
\bauthor{\bsnm{H{\'{e}}am}, \binits{P.}},
\bauthor{\bsnm{Kouchnarenko}, \binits{O.}},
\bauthor{\bsnm{Mantovani}, \binits{J.}},
\bauthor{\bsnm{M{\"{o}}dersheim}, \binits{S.}},
\bauthor{\bsnm{{von Oheimb}}, \binits{D.}},
\bauthor{\bsnm{Rusinowitch}, \binits{M.}},
\bauthor{\bsnm{Santiago}, \binits{J.}},
\bauthor{\bsnm{Turuani}, \binits{M.}},
\bauthor{\bsnm{Vigan{\`{o}}}, \binits{L.}},
\bauthor{\bsnm{Vigneron}, \binits{L.}}:
\bctitle{The {AVISPA} tool for the automated validation of internet security
  protocols and applications}.
In: \bbtitle{Proc. 17th Conference on Computer Aided Verification (CAV)}.
\bsertitle{LNCS},
vol. \bseriesno{3576},
pp. \bfpage{281}--\blpage{285}.
\bpublisher{Springer},
\blocation{Heidelberg}
(\byear{2005}).
\doiurl{10.1007/11513988\_27}
\end{bchapter}
\endbibitem

\bibitem[\protect\citeauthoryear{Sanzgiri et~al.}{2005}]{SanzgiriLDLSB05}
\begin{barticle}
\bauthor{\bsnm{Sanzgiri}, \binits{K.}},
\bauthor{\bsnm{LaFlamme}, \binits{D.}},
\bauthor{\bsnm{Dahill}, \binits{B.}},
\bauthor{\bsnm{Levine}, \binits{B.N.}},
\bauthor{\bsnm{Shields}, \binits{C.}},
\bauthor{\bsnm{Belding{-}Royer}, \binits{E.M.}}:
\batitle{Authenticated routing for ad hoc networks}.
\bjtitle{IEEE J. Sel. Areas Commun.}
\bvolume{23}(\bissue{3}),
\bfpage{598}--\blpage{610}
(\byear{2005})
\doiurl{10.1109/JSAC.2004.842547}
\end{barticle}
\endbibitem

\bibitem[\protect\citeauthoryear{Godskesen}{2006}]{Godskesen06}
\begin{bchapter}
\bauthor{\bsnm{Godskesen}, \binits{J.C.}}:
\bctitle{Formal verification of the {ARAN} protocol using the applied
  $\pi$-calculus}.
In: \bbtitle{Proc. 6th IFIP WG 1.7 Workshop on Issues in the Theory of Security
  (WITS)}
(\byear{2006}).
\bcomment{\url{http://www.itu.dk/~jcg/Papers/ARAN-APPpi.pdf}}
\end{bchapter}
\endbibitem

\bibitem[\protect\citeauthoryear{Blanchet}{2013}]{Blanchet13}
\begin{bchapter}
\bauthor{\bsnm{Blanchet}, \binits{B.}}:
\bctitle{Automatic verification of security protocols in the symbolic model:
  The verifier {ProVerif}}.
In: \bbtitle{Foundations of Security Analysis and Design {VII} (FOSAD),
  Tutorial Lectures}.
\bsertitle{LNCS},
vol. \bseriesno{8604},
pp. \bfpage{54}--\blpage{87}.
\bpublisher{Springer},
\blocation{Heidelberg}
(\byear{2013}).
\doiurl{10.1007/978-3-319-10082-1\_3}
\end{bchapter}
\endbibitem

\bibitem[\protect\citeauthoryear{Pura et~al.}{2009}]{PuraPB09}
\begin{bchapter}
\bauthor{\bsnm{Pura}, \binits{M.}},
\bauthor{\bsnm{Patriciu}, \binits{V.V.}},
\bauthor{\bsnm{Bica}, \binits{I.}}:
\bctitle{Simulation of an identity-based cryptography scheme for ad hoc
  networks}.
In: \bbtitle{Proc. 6th Conference on Security and Cryptography (SECRYPT)},
pp. \bfpage{135}--\blpage{139}.
\bpublisher{INSTICC Press},
\blocation{Lisbon}
(\byear{2009}).
\burl{https://www.naun.org/main/UPress/cc/19-311.pdf}
\end{bchapter}
\endbibitem

\bibitem[\protect\citeauthoryear{Pura and Buchs}{2015}]{PuraBuchs15}
\begin{barticle}
\bauthor{\bsnm{Pura}, \binits{M.}},
\bauthor{\bsnm{Buchs}, \binits{D.}}:
\batitle{Symbolic model checking of security protocols for ad hoc networks on
  any topologies}.
\bjtitle{Trans. Petri Nets Other Model. Concurr.}
\bvolume{10},
\bfpage{109}--\blpage{130}
(\byear{2015})
\doiurl{10.1007/978-3-662-48650-4\_6}
\end{barticle}
\endbibitem

\bibitem[\protect\citeauthoryear{Hostettler et~al.}{2011}]{HostettlerMLRB11}
\begin{barticle}
\bauthor{\bsnm{Hostettler}, \binits{S.}},
\bauthor{\bsnm{Marechal}, \binits{A.}},
\bauthor{\bsnm{Linard}, \binits{A.}},
\bauthor{\bsnm{Risoldi}, \binits{M.}},
\bauthor{\bsnm{Buchs}, \binits{D.}}:
\batitle{High-level {Petri} net model checking with {AlPiNA}}.
\bjtitle{Fundam. Informaticae}
\bvolume{113}(\bissue{3-4}),
\bfpage{229}--\blpage{264}
(\byear{2011})
\doiurl{10.3233/FI-2011-608}
\end{barticle}
\endbibitem

\bibitem[\protect\citeauthoryear{Sosnovich et~al.}{2013}]{SosnovichGN13}
\begin{bchapter}
\bauthor{\bsnm{Sosnovich}, \binits{A.}},
\bauthor{\bsnm{Grumberg}, \binits{O.}},
\bauthor{\bsnm{Nakibly}, \binits{G.}}:
\bctitle{Finding security vulnerabilities in a network protocol using
  parameterized systems}.
In: \bbtitle{Proc. 25th Conference on Computer Aided Verification (CAV)}.
\bsertitle{LNCS},
vol. \bseriesno{8044},
pp. \bfpage{724}--\blpage{739}.
\bpublisher{Springer},
\blocation{Heidelberg}
(\byear{2013}).
\doiurl{10.1007/978-3-642-39799-8\_51}
\end{bchapter}
\endbibitem

\bibitem[\protect\citeauthoryear{Clarke et~al.}{2004}]{ClarkeKL04}
\begin{bchapter}
\bauthor{\bsnm{Clarke}, \binits{E.M.}},
\bauthor{\bsnm{Kroening}, \binits{D.}},
\bauthor{\bsnm{Lerda}, \binits{F.}}:
\bctitle{A tool for checking {ANSI-C} programs}.
In: \bbtitle{Proc. 10th Conference on Tools and Algorithms for the Construction
  and Analysis of Systems (TACAS)}.
\bsertitle{LNCS},
vol. \bseriesno{2988},
pp. \bfpage{168}--\blpage{176}.
\bpublisher{Springer},
\blocation{Heidelberg}
(\byear{2004}).
\doiurl{10.1007/978-3-540-24730-2\_15}
\end{bchapter}
\endbibitem

\bibitem[\protect\citeauthoryear{Nakibly et~al.}{2012}]{NakiblyKGB12}
\begin{bchapter}
\bauthor{\bsnm{Nakibly}, \binits{G.}},
\bauthor{\bsnm{Kirshon}, \binits{A.}},
\bauthor{\bsnm{Gonikman}, \binits{D.}},
\bauthor{\bsnm{Boneh}, \binits{D.}}:
\bctitle{Persistent {OSPF} attacks}.
In: \bbtitle{Proc. 19th Network and Distributed System Security Symposium
  (NDSS)}.
\bpublisher{The Internet Society},
\blocation{Reston}
(\byear{2012}).
\burl{https://www.ndss-symposium.org/wp-content/uploads/2017/09/P01_3.pdf}
\end{bchapter}
\endbibitem

\bibitem[\protect\citeauthoryear{Nakibly et~al.}{2014}]{NakiblySMWE14}
\begin{bchapter}
\bauthor{\bsnm{Nakibly}, \binits{G.}},
\bauthor{\bsnm{Sosnovich}, \binits{A.}},
\bauthor{\bsnm{Menahem}, \binits{E.}},
\bauthor{\bsnm{Waizel}, \binits{A.}},
\bauthor{\bsnm{Elovici}, \binits{Y.}}:
\bctitle{{OSPF} vulnerability to persistent poisoning attacks: A systematic
  analysis}.
In: \bbtitle{Proc. 30th Annual Computer Security Applications Conference
  (ACSAC)},
pp. \bfpage{336}--\blpage{345}.
\bpublisher{ACM},
\blocation{New York}
(\byear{2014}).
\doiurl{10.1145/2664243.2664278}
\end{bchapter}
\endbibitem

\bibitem[\protect\citeauthoryear{Darville et~al.}{2022}]{DarvilleHIP22}
\begin{bchapter}
\bauthor{\bsnm{Darville}, \binits{C.}},
\bauthor{\bsnm{H{\"{o}}fner}, \binits{P.}},
\bauthor{\bsnm{Ivankovic}, \binits{F.}},
\bauthor{\bsnm{Pam}, \binits{A.}}:
\bctitle{Advanced models for the {OSPF} routing protocol}.
In: \bbtitle{Proc. 5th Workshop on Models for Formal Analysis of Real Systems
  (MARS@ETAPS)}.
\bsertitle{EPTCS},
vol. \bseriesno{355},
pp. \bfpage{13}--\blpage{26}
(\byear{2022}).
\doiurl{10.4204/EPTCS.355.2}
\end{bchapter}
\endbibitem

\bibitem[\protect\citeauthoryear{Andel et~al.}{2011}]{AndelBY11}
\begin{barticle}
\bauthor{\bsnm{Andel}, \binits{T.R.}},
\bauthor{\bsnm{Back}, \binits{G.}},
\bauthor{\bsnm{Yasinsac}, \binits{A.}}:
\batitle{Automating the security analysis process of secure ad hoc routing
  protocols}.
\bjtitle{Simul. Model. Pract. Theory}
\bvolume{19}(\bissue{9}),
\bfpage{2032}--\blpage{2049}
(\byear{2011})
\doiurl{10.1016/j.simpat.2011.05.008}
\end{barticle}
\endbibitem

\bibitem[\protect\citeauthoryear{Papadimitratos and
  Haas}{2002}]{PapadimitratosHaas02}
\begin{bchapter}
\bauthor{\bsnm{Papadimitratos}, \binits{P.}},
\bauthor{\bsnm{Haas}, \binits{Z.J.}}:
\bctitle{Secure routing for mobile ad hoc networks}.
In: \bbtitle{Proc. SCS Communication Networks and Distributed Systems Modeling
  and Simulation Conference (CNDS)},
pp. \bfpage{1}--\blpage{13}
(\byear{2002})
\end{bchapter}
\endbibitem

\bibitem[\protect\citeauthoryear{Cortier et~al.}{2012}]{CortierDD12}
\begin{bchapter}
\bauthor{\bsnm{Cortier}, \binits{V.}},
\bauthor{\bsnm{Degrieck}, \binits{J.}},
\bauthor{\bsnm{Delaune}, \binits{S.}}:
\bctitle{Analysing routing protocols: Four nodes topologies are sufficient}.
In: \bbtitle{Proc. 1st Conference on Principles of Security and Trust (POST)}.
\bsertitle{LNCS},
vol. \bseriesno{7215},
pp. \bfpage{30}--\blpage{50}.
\bpublisher{Springer},
\blocation{Heidelberg}
(\byear{2012}).
\doiurl{10.1007/978-3-642-28641-4\_3}
\end{bchapter}
\endbibitem

\bibitem[\protect\citeauthoryear{Berton et~al.}{2006}]{BertonYLM06}
\begin{bchapter}
\bauthor{\bsnm{Berton}, \binits{S.}},
\bauthor{\bsnm{Yin}, \binits{H.}},
\bauthor{\bsnm{Lin}, \binits{C.}},
\bauthor{\bsnm{Min}, \binits{G.}}:
\bctitle{Secure, disjoint, multipath source routing protocol ({SDMSR}) for
  mobile ad-hoc networks}.
In: \bbtitle{Proc. 5th Conference on Grid and Cooperative Computing (GCC)}.
\bpublisher{IEEE},
\blocation{New York}
(\byear{2006}).
\doiurl{10.1109/GCC.2006.86}
\end{bchapter}
\endbibitem

\bibitem[\protect\citeauthoryear{Nanz and Hankin}{2006}]{NanzHankin06}
\begin{barticle}
\bauthor{\bsnm{Nanz}, \binits{S.}},
\bauthor{\bsnm{Hankin}, \binits{C.}}:
\batitle{A framework for security analysis of mobile wireless networks}.
\bjtitle{Theor. Comput. Sci.}
\bvolume{367}(\bissue{1-2}),
\bfpage{203}--\blpage{227}
(\byear{2006})
\doiurl{10.1016/j.tcs.2006.08.036}
\end{barticle}
\endbibitem

\bibitem[\protect\citeauthoryear{Arnaud et~al.}{2014}]{ArnaudCD14}
\begin{barticle}
\bauthor{\bsnm{Arnaud}, \binits{M.}},
\bauthor{\bsnm{Cortier}, \binits{V.}},
\bauthor{\bsnm{Delaune}, \binits{S.}}:
\batitle{Modeling and verifying ad hoc routing protocols}.
\bjtitle{Inf. Comput.}
\bvolume{238},
\bfpage{30}--\blpage{67}
(\byear{2014})
\doiurl{10.1016/j.ic.2014.07.004}
\end{barticle}
\endbibitem

\bibitem[\protect\citeauthoryear{Godskesen}{2007}]{Godskesen07}
\begin{bchapter}
\bauthor{\bsnm{Godskesen}, \binits{J.C.}}:
\bctitle{A calculus for mobile ad hoc networks}.
In: \bbtitle{Proc. 9th Conference on Coordination Models and Languages
  (COORDINATION)}.
\bsertitle{LNCS},
vol. \bseriesno{4467},
pp. \bfpage{132}--\blpage{150}.
\bpublisher{Springer},
\blocation{Heidelberg}
(\byear{2007}).
\doiurl{10.1007/978-3-540-72794-1\_8}
\end{bchapter}
\endbibitem

\bibitem[\protect\citeauthoryear{Godskesen}{2008}]{Godskesen08}
\begin{bchapter}
\bauthor{\bsnm{Godskesen}, \binits{J.C.}}:
\bctitle{A calculus for mobile ad-hoc networks with static location binding}.
In: \bbtitle{Proc. 15th Workshop on Expressiveness in Concurrency (EXPRESS)}.
\bsertitle{ENTCS},
vol. \bseriesno{242},
pp. \bfpage{161}--\blpage{183}.
\bpublisher{Elsevier},
\blocation{Amsterdam}
(\byear{2008}).
\doiurl{10.1016/j.entcs.2009.06.018}
\end{bchapter}
\endbibitem

\bibitem[\protect\citeauthoryear{Merro}{2009}]{Merro09}
\begin{barticle}
\bauthor{\bsnm{Merro}, \binits{M.}}:
\batitle{An observational theory for mobile ad hoc networks (full version)}.
\bjtitle{Inf. Comput.}
\bvolume{207}(\bissue{2}),
\bfpage{194}--\blpage{208}
(\byear{2009})
\doiurl{10.1016/j.ic.2007.11.010}
\end{barticle}
\endbibitem

\bibitem[\protect\citeauthoryear{Merro and Sibilio}{2013}]{MerroSibilio13}
\begin{barticle}
\bauthor{\bsnm{Merro}, \binits{M.}},
\bauthor{\bsnm{Sibilio}, \binits{E.}}:
\batitle{A calculus of trustworthy ad hoc networks}.
\bjtitle{Formal Aspects Comput.}
\bvolume{25}(\bissue{5}),
\bfpage{801}--\blpage{832}
(\byear{2013})
\doiurl{10.1007/s00165-011-0210-7}
\end{barticle}
\endbibitem

\bibitem[\protect\citeauthoryear{Ghassemi et~al.}{2008}]{GhassemiFM08}
\begin{bchapter}
\bauthor{\bsnm{Ghassemi}, \binits{F.}},
\bauthor{\bsnm{Fokkink}, \binits{W.J.}},
\bauthor{\bsnm{Movaghar}, \binits{A.}}:
\bctitle{Restricted broadcast process theory}.
In: \bbtitle{Proc. 6th Conference on Software Engineering and Formal Methods
  (SEFM)},
pp. \bfpage{345}--\blpage{354}.
\bpublisher{IEEE},
\blocation{New York}
(\byear{2008}).
\doiurl{10.1109/SEFM.2008.25}
\end{bchapter}
\endbibitem

\bibitem[\protect\citeauthoryear{Ghassemi et~al.}{2010}]{GhassemiFM10}
\begin{barticle}
\bauthor{\bsnm{Ghassemi}, \binits{F.}},
\bauthor{\bsnm{Fokkink}, \binits{W.J.}},
\bauthor{\bsnm{Movaghar}, \binits{A.}}:
\batitle{Equational reasoning on mobile ad hoc networks}.
\bjtitle{Fundam. Informaticae}
\bvolume{105}(\bissue{4}),
\bfpage{375}--\blpage{415}
(\byear{2010})
\doiurl{10.3233/FI-2010-371}
\end{barticle}
\endbibitem

\bibitem[\protect\citeauthoryear{Ghassemi et~al.}{2011}]{GhassemiFM11}
\begin{barticle}
\bauthor{\bsnm{Ghassemi}, \binits{F.}},
\bauthor{\bsnm{Fokkink}, \binits{W.J.}},
\bauthor{\bsnm{Movaghar}, \binits{A.}}:
\batitle{Verification of mobile ad hoc networks: An algebraic approach}.
\bjtitle{Theor. Comput. Sci.}
\bvolume{412}(\bissue{28}),
\bfpage{3262}--\blpage{3282}
(\byear{2011})
\doiurl{10.1016/j.tcs.2011.03.017}
\end{barticle}
\endbibitem

\bibitem[\protect\citeauthoryear{Ghassemi and
  Fokkink}{2019}]{GhassemiFokkink19}
\begin{barticle}
\bauthor{\bsnm{Ghassemi}, \binits{F.}},
\bauthor{\bsnm{Fokkink}, \binits{W.J.}}:
\batitle{Reliable restricted process theory}.
\bjtitle{Fundam. Informaticae}
\bvolume{165}(\bissue{1}),
\bfpage{1}--\blpage{41}
(\byear{2019})
\doiurl{10.3233/FI-2019-1775}
\end{barticle}
\endbibitem

\bibitem[\protect\citeauthoryear{Kouzapas and
  Philippou}{2011}]{KouzapasPhilippou11}
\begin{bchapter}
\bauthor{\bsnm{Kouzapas}, \binits{D.}},
\bauthor{\bsnm{Philippou}, \binits{A.}}:
\bctitle{A process calculus for dynamic networks}.
In: \bbtitle{Proc. Joint 13th IFIP WG 6.1 Conference on Formal Methods for Open
  Object-Based Distributed Systems and 31st IFIP WG 6.1 Conference on Formal
  Techniques for Distributed Systems (FMOODS/FORTE)}.
\bsertitle{LNCS},
vol. \bseriesno{6722},
pp. \bfpage{213}--\blpage{227}.
\bpublisher{Springer},
\blocation{Heidelberg}
(\byear{2011}).
\doiurl{10.1007/978-3-642-21461-5\_14}
\end{bchapter}
\endbibitem

\bibitem[\protect\citeauthoryear{Vasudevan et~al.}{2004}]{VasudevanKT04}
\begin{bchapter}
\bauthor{\bsnm{Vasudevan}, \binits{S.}},
\bauthor{\bsnm{Kurose}, \binits{J.F.}},
\bauthor{\bsnm{Towsley}, \binits{D.F.}}:
\bctitle{Design and analysis of a leader election algorithm for mobile ad hoc
  networks}.
In: \bbtitle{Proc. 12th Conference on Network Protocols (ICNP)},
pp. \bfpage{350}--\blpage{360}.
\bpublisher{IEEE},
\blocation{New York}
(\byear{2004}).
\doiurl{10.1109/ICNP.2004.1348124}
\end{bchapter}
\endbibitem

\bibitem[\protect\citeauthoryear{Ghassemi et~al.}{2011}]{GhassemiTMF11}
\begin{bchapter}
\bauthor{\bsnm{Ghassemi}, \binits{F.}},
\bauthor{\bsnm{Talebi}, \binits{M.}},
\bauthor{\bsnm{Movaghar}, \binits{A.}},
\bauthor{\bsnm{Fokkink}, \binits{W.J.}}:
\bctitle{Stochastic restricted broadcast process theory}.
In: \bbtitle{Proc. 8th European Performance Engineering Workshop (EPEW)}.
\bsertitle{LNCS},
vol. \bseriesno{6977},
pp. \bfpage{72}--\blpage{86}.
\bpublisher{Springer},
\blocation{Heidelberg}
(\byear{2011}).
\doiurl{10.1007/978-3-642-24749-1\_7}
\end{bchapter}
\endbibitem

\bibitem[\protect\citeauthoryear{Fehnker et~al.}{2012a}]{FehnkerGHMPT12a}
\begin{bchapter}
\bauthor{\bsnm{Fehnker}, \binits{A.}},
\bauthor{\bsnm{{van Glabbeek}}, \binits{R.J.}},
\bauthor{\bsnm{H{\"{o}}fner}, \binits{P.}},
\bauthor{\bsnm{McIver}, \binits{A.}},
\bauthor{\bsnm{Portmann}, \binits{M.}},
\bauthor{\bsnm{Tan}, \binits{W.L.}}:
\bctitle{A process algebra for wireless mesh networks}.
In: \bbtitle{Proc. 21st European Symposium on Programming (ESOP)}.
\bsertitle{LNCS},
vol. \bseriesno{7211},
pp. \bfpage{295}--\blpage{315}.
\bpublisher{Springer},
\blocation{Heidelberg}
(\byear{2012}).
\doiurl{10.1007/978-3-642-28869-2\_15}
\end{bchapter}
\endbibitem

\bibitem[\protect\citeauthoryear{Fehnker et~al.}{2012b}]{FehnkerGHMPT12b}
\begin{bchapter}
\bauthor{\bsnm{Fehnker}, \binits{A.}},
\bauthor{\bsnm{{van Glabbeek}}, \binits{R.J.}},
\bauthor{\bsnm{H{\"{o}}fner}, \binits{P.}},
\bauthor{\bsnm{McIver}, \binits{A.}},
\bauthor{\bsnm{Portmann}, \binits{M.}},
\bauthor{\bsnm{Tan}, \binits{W.L.}}:
\bctitle{Automated analysis of {AODV} using {UPPAAL}}.
In: \bbtitle{Proc. 18th Conference on Tools and Algorithms for the Construction
  and Analysis of Systems (TACAS)}.
\bsertitle{LNCS},
vol. \bseriesno{7214},
pp. \bfpage{173}--\blpage{187}.
\bpublisher{Springer},
\blocation{Heidelberg}
(\byear{2012}).
\doiurl{10.1007/978-3-642-28756-5\_13}
\end{bchapter}
\endbibitem

\bibitem[\protect\citeauthoryear{H{\"{o}}fner et~al.}{2012}]{HofnerGTPMF12}
\begin{bchapter}
\bauthor{\bsnm{H{\"{o}}fner}, \binits{P.}},
\bauthor{\bsnm{{van Glabbeek}}, \binits{R.J.}},
\bauthor{\bsnm{Tan}, \binits{W.L.}},
\bauthor{\bsnm{Portmann}, \binits{M.}},
\bauthor{\bsnm{McIver}, \binits{A.}},
\bauthor{\bsnm{Fehnker}, \binits{A.}}:
\bctitle{A rigorous analysis of {AODV} and its variants}.
In: \bbtitle{Proc. 15th Conference on Modeling, Analysis and Simulation of
  Wireless and Mobile Systems (MSWiM)},
pp. \bfpage{203}--\blpage{212}.
\bpublisher{ACM},
\blocation{New York}
(\byear{2012}).
\doiurl{10.1145/2387238.2387274}
\end{bchapter}
\endbibitem

\bibitem[\protect\citeauthoryear{{van Glabbeek} et~al.}{2016}]{GlabbeekHPT16}
\begin{barticle}
\bauthor{\bsnm{{van Glabbeek}}, \binits{R.J.}},
\bauthor{\bsnm{H{\"{o}}fner}, \binits{P.}},
\bauthor{\bsnm{Portmann}, \binits{M.}},
\bauthor{\bsnm{Tan}, \binits{W.L.}}:
\batitle{Modelling and verifying the {AODV} routing protocol}.
\bjtitle{Distributed Comput.}
\bvolume{29}(\bissue{4}),
\bfpage{279}--\blpage{315}
(\byear{2016})
\doiurl{10.1007/s00446-015-0262-7}
\end{barticle}
\endbibitem

\bibitem[\protect\citeauthoryear{Bourke et~al.}{2014}]{BourkeGH14}
\begin{bchapter}
\bauthor{\bsnm{Bourke}, \binits{T.}},
\bauthor{\bsnm{{van Glabbeek}}, \binits{R.J.}},
\bauthor{\bsnm{H{\"{o}}fner}, \binits{P.}}:
\bctitle{A mechanized proof of loop freedom of the (untimed) {AODV} routing
  protocol}.
In: \bbtitle{Proc. 12th Symposium on Automated Technology for Verification and
  Analysis (ATVA)}.
\bsertitle{LNCS},
vol. \bseriesno{8837},
pp. \bfpage{47}--\blpage{63}.
\bpublisher{Springer},
\blocation{Heidelberg}
(\byear{2014}).
\doiurl{10.1007/978-3-319-11936-6\_5}
\end{bchapter}
\endbibitem

\bibitem[\protect\citeauthoryear{Paulson}{1994}]{Paulson94}
\begin{bbook}
\bauthor{\bsnm{Paulson}, \binits{L.C.}}:
\bbtitle{Isabelle - {A} Generic Theorem Prover}.
\bsertitle{LNCS},
vol. \bseriesno{828}.
\bpublisher{Springer},
\blocation{Heidelberg}
(\byear{1994}).
\doiurl{10.1007/BFb0030541}
\end{bbook}
\endbibitem

\bibitem[\protect\citeauthoryear{Bourke et~al.}{2016}]{BourkeGH16}
\begin{barticle}
\bauthor{\bsnm{Bourke}, \binits{T.}},
\bauthor{\bsnm{{van Glabbeek}}, \binits{R.J.}},
\bauthor{\bsnm{H{\"{o}}fner}, \binits{P.}}:
\batitle{Mechanizing a process algebra for network protocols}.
\bjtitle{J. Autom. Reason.}
\bvolume{56}(\bissue{3}),
\bfpage{309}--\blpage{341}
(\byear{2016})
\doiurl{10.1007/s10817-015-9358-9}
\end{barticle}
\endbibitem

\bibitem[\protect\citeauthoryear{{van Glabbeek} et~al.}{2018}]{GlabbeekHW18}
\begin{bchapter}
\bauthor{\bsnm{{van Glabbeek}}, \binits{R.J.}},
\bauthor{\bsnm{H{\"{o}}fner}, \binits{P.}},
\bauthor{\bsnm{{van der Wal}}, \binits{D.}}:
\bctitle{Analysing {AWN}-specifications using {mCRL2} (extended abstract)}.
In: \bbtitle{Proc. 14th Conference on Integrated Formal Methods (IFM)}.
\bsertitle{LNCS},
vol. \bseriesno{11023},
pp. \bfpage{398}--\blpage{418}.
\bpublisher{Springer},
\blocation{Heidelberg}
(\byear{2018}).
\doiurl{10.1007/978-3-319-98938-9\_23}
\end{bchapter}
\endbibitem

\bibitem[\protect\citeauthoryear{Bres et~al.}{2016}]{BresGH16}
\begin{bchapter}
\bauthor{\bsnm{Bres}, \binits{E.}},
\bauthor{\bsnm{{van Glabbeek}}, \binits{R.J.}},
\bauthor{\bsnm{H{\"{o}}fner}, \binits{P.}}:
\bctitle{A timed process algebra for wireless networks with an application in
  routing (extended abstract)}.
In: \bbtitle{Proc. 25th European Symposium on Programming (ESOP)}.
\bsertitle{LNCS},
vol. \bseriesno{9632},
pp. \bfpage{95}--\blpage{122}.
\bpublisher{Springer},
\blocation{Heidelberg}
(\byear{2016}).
\doiurl{10.1007/978-3-662-49498-1\_5}
\end{bchapter}
\endbibitem

\bibitem[\protect\citeauthoryear{Barry et~al.}{2020}]{BarryGH20}
\begin{bchapter}
\bauthor{\bsnm{Barry}, \binits{R.}},
\bauthor{\bsnm{{van Glabbeek}}, \binits{R.J.}},
\bauthor{\bsnm{H{\"{o}}fner}, \binits{P.}}:
\bctitle{Formalising the optimised link state routing protocol}.
In: \bbtitle{Proc. 4th Workshop on Models for Formal Analysis of Real Systems
  (MARS@ETAPS)}.
\bsertitle{EPTCS},
vol. \bseriesno{316},
pp. \bfpage{40}--\blpage{71}
(\byear{2020}).
\doiurl{10.4204/EPTCS.316.3}
\end{bchapter}
\endbibitem

\bibitem[\protect\citeauthoryear{Drury et~al.}{2020}]{DruryHW20}
\begin{bchapter}
\bauthor{\bsnm{Drury}, \binits{J.}},
\bauthor{\bsnm{H{\"{o}}fner}, \binits{P.}},
\bauthor{\bsnm{Wang}, \binits{W.}}:
\bctitle{Formal models of the {OSPF} routing protocol}.
In: \bbtitle{Proc. 4th Workshop on Models for Formal Analysis of Real Systems
  (MARS@ETAPS)}.
\bsertitle{EPTCS},
vol. \bseriesno{316},
pp. \bfpage{72}--\blpage{120}
(\byear{2020}).
\doiurl{10.4204/EPTCS.316.4}
\end{bchapter}
\endbibitem

\bibitem[\protect\citeauthoryear{Lanese and
  Sangiorgi}{2010}]{LaneseSangiorgi10}
\begin{barticle}
\bauthor{\bsnm{Lanese}, \binits{I.}},
\bauthor{\bsnm{Sangiorgi}, \binits{D.}}:
\batitle{An operational semantics for a calculus for wireless systems}.
\bjtitle{Theor. Comput. Sci.}
\bvolume{411}(\bissue{19}),
\bfpage{1928}--\blpage{1948}
(\byear{2010})
\doiurl{10.1016/j.tcs.2010.01.023}
\end{barticle}
\endbibitem

\bibitem[\protect\citeauthoryear{Wang and Lu}{2012}]{WangLu12}
\begin{bchapter}
\bauthor{\bsnm{Wang}, \binits{M.}},
\bauthor{\bsnm{Lu}, \binits{Y.}}:
\bctitle{A timed calculus for mobile ad hoc networks}.
In: \bbtitle{Proc. 1st Workshop on Formal Techniques for Safety-Critical
  Systems (FTSCS)}.
\bsertitle{EPTCS},
vol. \bseriesno{105},
pp. \bfpage{118}--\blpage{134}
(\byear{2012}).
\doiurl{10.4204/EPTCS.105.9}
\end{bchapter}
\endbibitem

\bibitem[\protect\citeauthoryear{Chen and Zhu}{2023}]{ChenZhu23}
\begin{barticle}
\bauthor{\bsnm{Chen}, \binits{N.}},
\bauthor{\bsnm{Zhu}, \binits{H.}}:
\batitle{{IoT} modeling and verification: From the {CaIT} calculus to {{\sc
  Uppaal}}}.
\bjtitle{{IEICE} Trans. Inf. Syst.}
\bvolume{106}(\bissue{9}),
\bfpage{1507}--\blpage{1518}
(\byear{2023})
\doiurl{10.1587/TRANSINF.2022EDP7223}
\end{barticle}
\endbibitem

\bibitem[\protect\citeauthoryear{Lanotte and Merro}{2018}]{LanotteMerro18}
\begin{barticle}
\bauthor{\bsnm{Lanotte}, \binits{R.}},
\bauthor{\bsnm{Merro}, \binits{M.}}:
\batitle{A semantic theory of the {Internet of Things}}.
\bjtitle{Inf. Comput.}
\bvolume{259}(\bissue{1}),
\bfpage{72}--\blpage{101}
(\byear{2018})
\doiurl{10.1016/J.IC.2018.01.001}
\end{barticle}
\endbibitem

\bibitem[\protect\citeauthoryear{Milner et~al.}{1992}]{MilnerPW92}
\begin{barticle}
\bauthor{\bsnm{Milner}, \binits{R.}},
\bauthor{\bsnm{Parrow}, \binits{J.}},
\bauthor{\bsnm{Walker}, \binits{D.}}:
\batitle{A calculus of mobile processes, {I + II}}.
\bjtitle{Inf. Comput.}
\bvolume{100}(\bissue{1}),
\bfpage{1}--\blpage{77}
(\byear{1992})
\doiurl{10.1016/0890-5401(92)90008-4}
\end{barticle}
\endbibitem

\bibitem[\protect\citeauthoryear{Cardelli and Gordon}{2000}]{CardelliGordon00}
\begin{barticle}
\bauthor{\bsnm{Cardelli}, \binits{L.}},
\bauthor{\bsnm{Gordon}, \binits{A.D.}}:
\batitle{Mobile ambients}.
\bjtitle{Theor. Comput. Sci.}
\bvolume{240}(\bissue{1}),
\bfpage{177}--\blpage{213}
(\byear{2000})
\doiurl{10.1016/S0304-3975(99)00231-5}
\end{barticle}
\endbibitem

\bibitem[\protect\citeauthoryear{Sewell et~al.}{1998}]{SewellWP98}
\begin{bchapter}
\bauthor{\bsnm{Sewell}, \binits{P.}},
\bauthor{\bsnm{Wojciechowski}, \binits{P.T.}},
\bauthor{\bsnm{Pierce}, \binits{B.C.}}:
\bctitle{Location-independent communication for mobile agents: {A} two-level
  architecture}.
In: \bbtitle{Proc. ICCL Workshop on Internet Programming Languages}.
\bsertitle{LNCS},
vol. \bseriesno{1686},
pp. \bfpage{1}--\blpage{31}.
\bpublisher{Springer},
\blocation{Heidelberg}
(\byear{1998}).
\doiurl{10.1007/3-540-47959-7\_1}
\end{bchapter}
\endbibitem

\bibitem[\protect\citeauthoryear{Sewell et~al.}{2010}]{SewellWU10}
\begin{barticle}
\bauthor{\bsnm{Sewell}, \binits{P.}},
\bauthor{\bsnm{Wojciechowski}, \binits{P.T.}},
\bauthor{\bsnm{Unyapoth}, \binits{A.}}:
\batitle{Nomadic {Pict}: Programming languages, communication infrastructure
  overlays, and semantics for mobile computation}.
\bjtitle{ACM Trans. Program. Lang. Syst.}
\bvolume{32}(\bissue{4}),
\bfpage{12}--\blpage{11263}
(\byear{2010})
\doiurl{10.1145/1734206.1734209}
\end{barticle}
\endbibitem

\bibitem[\protect\citeauthoryear{Ene and Muntean}{2001}]{EneMuntean01}
\begin{bchapter}
\bauthor{\bsnm{Ene}, \binits{C.}},
\bauthor{\bsnm{Muntean}, \binits{T.}}:
\bctitle{A broadcast-based calculus for communicating systems}.
In: \bbtitle{Proc. 6th IPDPS Workshop on Formal Methods for Parallel
  Programming: Theory and Applications (FMPPTA)}.
\bpublisher{IEEE},
\blocation{New York}
(\byear{2001}).
\doiurl{10.1109/IPDPS.2001.925136}
\end{bchapter}
\endbibitem

\bibitem[\protect\citeauthoryear{Prasad}{2006}]{Prasad06}
\begin{bchapter}
\bauthor{\bsnm{Prasad}, \binits{K.V.S.}}:
\bctitle{A prospectus for mobile broadcasting systems}.
In: \bbtitle{Proc. Workshop "Essays on Algebraic Process Calculi" (APC 25)}.
\bsertitle{ENTCS},
vol. \bseriesno{162},
pp. \bfpage{295}--\blpage{300}.
\bpublisher{Elsevier},
\blocation{Amsterdam}
(\byear{2006}).
\doiurl{10.1016/j.entcs.2005.12.096}
\end{bchapter}
\endbibitem

\bibitem[\protect\citeauthoryear{Singh et~al.}{2010}]{SinghRS10}
\begin{barticle}
\bauthor{\bsnm{Singh}, \binits{A.}},
\bauthor{\bsnm{Ramakrishnan}, \binits{C.R.}},
\bauthor{\bsnm{Smolka}, \binits{S.A.}}:
\batitle{A process calculus for mobile ad hoc networks}.
\bjtitle{Sci. Comput. Program.}
\bvolume{75}(\bissue{6}),
\bfpage{440}--\blpage{469}
(\byear{2010})
\doiurl{10.1016/j.scico.2009.07.008}
\end{barticle}
\endbibitem

\bibitem[\protect\citeauthoryear{Bengtson et~al.}{2011}]{BengtsonJPV11}
\begin{botherref}
\oauthor{\bsnm{Bengtson}, \binits{J.}},
\oauthor{\bsnm{Johansson}, \binits{M.}},
\oauthor{\bsnm{Parrow}, \binits{J.}},
\oauthor{\bsnm{Victor}, \binits{B.}}:
Psi-calculi: A framework for mobile processes with nominal data and logic.
Log. Methods Comput. Sci.
\textbf{7}(1)
(2011)
\doiurl{10.2168/LMCS-7(1:11)2011}
\end{botherref}
\endbibitem

\bibitem[\protect\citeauthoryear{Borgstr{\"{o}}m
  et~al.}{2011}]{BorgstromHJRVPP11}
\begin{bchapter}
\bauthor{\bsnm{Borgstr{\"{o}}m}, \binits{J.}},
\bauthor{\bsnm{Huang}, \binits{S.}},
\bauthor{\bsnm{Johansson}, \binits{M.}},
\bauthor{\bsnm{Raabjerg}, \binits{P.}},
\bauthor{\bsnm{Victor}, \binits{B.}},
\bauthor{\bsnm{Pohjola}, \binits{J.{\AA}.}},
\bauthor{\bsnm{Parrow}, \binits{J.}}:
\bctitle{Broadcast psi-calculi with an application to wireless protocols}.
In: \bbtitle{Proc. 9th Conference on Software Engineering and Formal Methods
  (SEFM)}.
\bsertitle{LNCS},
vol. \bseriesno{7041},
pp. \bfpage{74}--\blpage{89}.
\bpublisher{Springer},
\blocation{Heidelberg}
(\byear{2011}).
\doiurl{10.1007/978-3-642-24690-6\_7}
\end{bchapter}
\endbibitem

\bibitem[\protect\citeauthoryear{Godskesen et~al.}{2009}]{GodskesenHK09}
\begin{bchapter}
\bauthor{\bsnm{Godskesen}, \binits{J.C.}},
\bauthor{\bsnm{H{\"{u}}ttel}, \binits{H.}},
\bauthor{\bsnm{K{\"{u}}hnrich}, \binits{M.}}:
\bctitle{Verification of correspondence assertions in a calculus for mobile ad
  hoc networks}.
In: \bbtitle{Proc. 7th Workshop on the Foundations of Coordination Languages
  and Software Architectures (FOCLASA)}.
\bsertitle{ENTCS},
vol. \bseriesno{229},
pp. \bfpage{77}--\blpage{93}.
\bpublisher{Elsevier},
\blocation{Amsterdam}
(\byear{2009}).
\doiurl{10.1016/j.entcs.2009.06.030}
\end{bchapter}
\endbibitem

\bibitem[\protect\citeauthoryear{Hennessy}{2007}]{Hennessy07}
\begin{bbook}
\bauthor{\bsnm{Hennessy}, \binits{M.C.B.}}:
\bbtitle{A Distributed Pi-calculus}.
\bpublisher{Cambridge University Press},
\blocation{Cambridge}
(\byear{2007}).
\doiurl{10.1017/CBO9780511611063}
\end{bbook}
\endbibitem

\bibitem[\protect\citeauthoryear{Abadi and Gordon}{1997}]{AbadiGordon97}
\begin{bchapter}
\bauthor{\bsnm{Abadi}, \binits{M.}},
\bauthor{\bsnm{Gordon}, \binits{A.D.}}:
\bctitle{A calculus for cryptographic protocols: The spi calculus}.
In: \bbtitle{Proc. 4th Conference on Computer and Communications Security
  (CCS)},
pp. \bfpage{36}--\blpage{47}.
\bpublisher{ACM},
\blocation{New York}
(\byear{1997}).
\doiurl{10.1145/266420.266432}
\end{bchapter}
\endbibitem

\bibitem[\protect\citeauthoryear{Perkins}{1996}]{Perkins96}
\begin{botherref}
\oauthor{\bsnm{Perkins}, \binits{C.E.}}:
IP Mobility Support. RFC 2002.
IETF,
(1996).
IETF.
\url{https://datatracker.ietf.org/doc/html/rfc2002}
\end{botherref}
\endbibitem

\bibitem[\protect\citeauthoryear{Godskesen}{2010}]{Godskesen10}
\begin{bchapter}
\bauthor{\bsnm{Godskesen}, \binits{J.C.}}:
\bctitle{Observables for mobile and wireless broadcasting systems}.
In: \bbtitle{Proc. 12th Conference on Coordination Models and Languages
  (COORDINATION)}.
\bsertitle{LNCS},
vol. \bseriesno{6116},
pp. \bfpage{1}--\blpage{15}.
\bpublisher{Springer},
\blocation{Heidelberg}
(\byear{2010}).
\doiurl{10.1007/978-3-642-13414-2\_1}
\end{bchapter}
\endbibitem

\bibitem[\protect\citeauthoryear{Abadi and Fournet}{2001}]{AbadiFournet01}
\begin{bchapter}
\bauthor{\bsnm{Abadi}, \binits{M.}},
\bauthor{\bsnm{Fournet}, \binits{C.}}:
\bctitle{Mobile values, new names, and secure communication}.
In: \bbtitle{Proc. 28th Symposium on Principles of Programming Languages
  (POPL)},
pp. \bfpage{104}--\blpage{115}.
\bpublisher{ACM},
\blocation{New York}
(\byear{2001}).
\doiurl{10.1145/360204.360213}
\end{bchapter}
\endbibitem

\bibitem[\protect\citeauthoryear{Chr{\'{e}}tien and
  Delaune}{2013}]{ChretienDelaune13}
\begin{bchapter}
\bauthor{\bsnm{Chr{\'{e}}tien}, \binits{R.}},
\bauthor{\bsnm{Delaune}, \binits{S.}}:
\bctitle{Formal analysis of privacy for routing protocols in mobile ad hoc
  networks}.
In: \bbtitle{Proc. 2nd Conference on Principles Of Security and Trust (POST)}.
\bsertitle{LNCS},
vol. \bseriesno{7796},
pp. \bfpage{1}--\blpage{20}.
\bpublisher{Springer},
\blocation{Heidelberg}
(\byear{2013}).
\doiurl{10.1007/978-3-642-36830-1\_1}
\end{bchapter}
\endbibitem

\bibitem[\protect\citeauthoryear{Kong and Hong}{2003}]{KongHong03}
\begin{bchapter}
\bauthor{\bsnm{Kong}, \binits{J.}},
\bauthor{\bsnm{Hong}, \binits{X.}}:
\bctitle{{ANODR}: Anonymous on demand routing with untraceable routes for
  mobile ad-hoc networks}.
In: \bbtitle{Proc. 4th Symposium on Mobile Ad Hoc Networking and Computing
  (MobiHoc)},
pp. \bfpage{291}--\blpage{302}.
\bpublisher{ACM},
\blocation{New York}
(\byear{2003}).
\doiurl{10.1145/778415.778449}
\end{bchapter}
\endbibitem

\bibitem[\protect\citeauthoryear{Prasad}{1995}]{Prasad95}
\begin{barticle}
\bauthor{\bsnm{Prasad}, \binits{K.V.S.}}:
\batitle{A calculus of broadcasting systems}.
\bjtitle{Sci. Comput. Program.}
\bvolume{25}(\bissue{2-3}),
\bfpage{285}--\blpage{327}
(\byear{1995})
\doiurl{10.1016/0167-6423(95)00017-8}
\end{barticle}
\endbibitem

\bibitem[\protect\citeauthoryear{Roman et~al.}{1997}]{RomanMP97}
\begin{barticle}
\bauthor{\bsnm{Roman}, \binits{G.}},
\bauthor{\bsnm{McCann}, \binits{P.J.}},
\bauthor{\bsnm{Plun}, \binits{J.Y.}}:
\batitle{Mobile {UNITY}: Reasoning and specification in mobile computing}.
\bjtitle{ACM Trans. Softw. Eng. Methodol.}
\bvolume{6}(\bissue{3}),
\bfpage{250}--\blpage{282}
(\byear{1997})
\doiurl{10.1145/258077.258079}
\end{barticle}
\endbibitem

\bibitem[\protect\citeauthoryear{Chandy and Misra}{1988}]{ChandyMisra88}
\begin{bbook}
\bauthor{\bsnm{Chandy}, \binits{K.M.}},
\bauthor{\bsnm{Misra}, \binits{J.}}:
\bbtitle{Parallel Program Design: A Foundation}.
\bpublisher{Addison-Wesley},
\blocation{Boston}
(\byear{1988})
\end{bbook}
\endbibitem

\bibitem[\protect\citeauthoryear{Smith}{2000}]{Smith00}
\begin{bbook}
\bauthor{\bsnm{Smith}, \binits{G.}}:
\bbtitle{The Object-Z Specification Language}.
\bsertitle{Advances in Formal Methods}.
\bpublisher{Kluwer},
\blocation{Alphen aan den Rijn}
(\byear{2000}).
\doiurl{10.1007/978-1-4615-5265-9}
\end{bbook}
\endbibitem

\bibitem[\protect\citeauthoryear{Smith}{2004}]{Smith04}
\begin{bchapter}
\bauthor{\bsnm{Smith}, \binits{G.}}:
\bctitle{A framework for modelling and analysing mobile systems}.
In: \bbtitle{Proc. 27th Australasian Computer Science Conference (ACSC2004)}.
\bsertitle{CRPIT},
vol. \bseriesno{26},
pp. \bfpage{193}--\blpage{202}.
\bpublisher{Australian Computer Society},
\blocation{Sydney}
(\byear{2004}).
\burl{http://crpit.scem.westernsydney.edu.au/abstracts/CRPITV26Smith.html}
\end{bchapter}
\endbibitem

\bibitem[\protect\citeauthoryear{Wu et~al.}{2013}]{WuSZ13}
\begin{bchapter}
\bauthor{\bsnm{Wu}, \binits{X.}},
\bauthor{\bsnm{Sanders}, \binits{J.W.}},
\bauthor{\bsnm{Zhu}, \binits{H.}}:
\bctitle{Formal modelling and analysis of {AODV}}.
In: \bbtitle{Proc. 18th Conference on Engineering of Complex Computer Systems
  (ICECCS)},
pp. \bfpage{93}--\blpage{100}.
\bpublisher{IEEE},
\blocation{New York}
(\byear{2013}).
\doiurl{10.1109/ICECCS.2013.22}
\end{bchapter}
\endbibitem

\bibitem[\protect\citeauthoryear{Kamali and Petre}{2016}]{KamaliPetre16}
\begin{bchapter}
\bauthor{\bsnm{Kamali}, \binits{M.}},
\bauthor{\bsnm{Petre}, \binits{L.}}:
\bctitle{Modelling link state routing in {Event-B}}.
In: \bbtitle{Proc. 21st Conference on Engineering of Complex Computer Systems
  (ICECCS)},
pp. \bfpage{207}--\blpage{210}.
\bpublisher{IEEE},
\blocation{New York}
(\byear{2016}).
\doiurl{10.1109/ICECCS.2016.035}
\end{bchapter}
\endbibitem

\bibitem[\protect\citeauthoryear{Abrial}{2010}]{Abrial10}
\begin{bbook}
\bauthor{\bsnm{Abrial}, \binits{J.-R.}}:
\bbtitle{Modeling in Event-B: System an Software Design}.
\bpublisher{Cambridge University Press},
\blocation{Cambridge}
(\byear{2010}).
\doiurl{10.1017/CBO9781139195881}
\end{bbook}
\endbibitem

\bibitem[\protect\citeauthoryear{Kamali and Petre}{2017}]{KamaliPetre17}
\begin{bchapter}
\bauthor{\bsnm{Kamali}, \binits{M.}},
\bauthor{\bsnm{Petre}, \binits{L.}}:
\bctitle{{{\sc Uppaal}} vs {Event-B} for modelling optimised link state
  routing}.
In: \bbtitle{Proc. 11th Conference on Verification and Evaluation of Computer
  and Communication Systems (VECoS)}.
\bsertitle{LNCS},
vol. \bseriesno{10466},
pp. \bfpage{189}--\blpage{203}.
\bpublisher{Springer},
\blocation{Heidelberg}
(\byear{2017}).
\doiurl{10.1007/978-3-319-66176-6\_13}
\end{bchapter}
\endbibitem

\bibitem[\protect\citeauthoryear{Fakhfakh et~al.}{2016}]{FakhfakhTKM16}
\begin{bchapter}
\bauthor{\bsnm{Fakhfakh}, \binits{F.}},
\bauthor{\bsnm{Tounsi}, \binits{M.}},
\bauthor{\bsnm{Kacem}, \binits{A.H.}},
\bauthor{\bsnm{Mosbah}, \binits{M.}}:
\bctitle{Towards a formal model for dynamic networks through refinement and
  evolving graphs}.
In: \bbtitle{Revised Selected Papers from Software Engineering, Artificial
  Intelligence, Networking and Parallel/Distributed Computing (SNDP)}.
\bsertitle{Studies in Computational Intelligence},
vol. \bseriesno{612},
pp. \bfpage{227}--\blpage{243}.
\bpublisher{Springer},
\blocation{Heidelberg}
(\byear{2016}).
\doiurl{10.1007/978-3-319-23509-7\_16}
\end{bchapter}
\endbibitem

\bibitem[\protect\citeauthoryear{Ferreira}{2004}]{Ferreira04}
\begin{barticle}
\bauthor{\bsnm{Ferreira}, \binits{A.}}:
\batitle{Building a reference combinatorial model for {MANETs}}.
\bjtitle{IEEE Netw.}
\bvolume{18}(\bissue{5}),
\bfpage{24}--\blpage{29}
(\byear{2004})
\doiurl{10.1109/MNET.2004.1337732}
\end{barticle}
\endbibitem

\bibitem[\protect\citeauthoryear{Gurevich}{1993}]{Gurevich93}
\begin{bchapter}
\bauthor{\bsnm{Gurevich}, \binits{Y.}}:
\bctitle{Evolving algebras 1993: Lipari guide}.
In: \beditor{\bsnm{B{\"{o}}rger}, \binits{E.}} (ed.)
\bbtitle{Specification and Validation Methods},
pp. \bfpage{9}--\blpage{36}.
\bpublisher{Oxford University Press},
\blocation{Oxford}
(\byear{1993}).
\burl{https://www.researchgate.net/publication/2241959_Evolving_Algebras_1993_Lipari_Guide}
\end{bchapter}
\endbibitem

\bibitem[\protect\citeauthoryear{Bencz{\'{u}}r et~al.}{2003}]{BenczurGL03}
\begin{bchapter}
\bauthor{\bsnm{Bencz{\'{u}}r}, \binits{A.A.}},
\bauthor{\bsnm{Gl{\"{a}}sser}, \binits{U.}},
\bauthor{\bsnm{Lukovszki}, \binits{T.}}:
\bctitle{Formal description of a distributed location service for mobile ad hoc
  networks}.
In: \bbtitle{Proc. 10th Workshop on Abstract State Machines, Advances in Theory
  and Practice (ASM)}.
\bsertitle{LNCS},
vol. \bseriesno{2589},
pp. \bfpage{204}--\blpage{217}.
\bpublisher{Springer},
\blocation{Heidelberg}
(\byear{2003}).
\doiurl{10.1007/3-540-36498-6\_11}
\end{bchapter}
\endbibitem

\bibitem[\protect\citeauthoryear{Gl{\"{a}}sser and Gu}{2005}]{GlasserGu05}
\begin{barticle}
\bauthor{\bsnm{Gl{\"{a}}sser}, \binits{U.}},
\bauthor{\bsnm{Gu}, \binits{Q.}}:
\batitle{Formal description and analysis of a distributed location service for
  mobile ad hoc networks}.
\bjtitle{Theor. Comput. Sci.}
\bvolume{336}(\bissue{2-3}),
\bfpage{285}--\blpage{309}
(\byear{2005})
\doiurl{10.1016/j.tcs.2004.11.009}
\end{barticle}
\endbibitem

\bibitem[\protect\citeauthoryear{Gl{\"{a}}sser and
  Prinz}{2005}]{GlasserPrinz05}
\begin{bchapter}
\bauthor{\bsnm{Gl{\"{a}}sser}, \binits{U.}},
\bauthor{\bsnm{Prinz}, \binits{A.}}:
\bctitle{{ASM} and {SDL} models of geographic routing in mobile ad hoc
  networks}.
In: \bbtitle{Proc. 12th SDL Forum}.
\bsertitle{LNCS},
vol. \bseriesno{3530},
pp. \bfpage{162}--\blpage{173}.
\bpublisher{Springer},
\blocation{Heidelberg}
(\byear{2005}).
\doiurl{10.1007/11506843\_11}
\end{bchapter}
\endbibitem

\bibitem[\protect\citeauthoryear{Rockstr{\"{o}}m and
  Saracco}{1982}]{RockstromSaracco82}
\begin{barticle}
\bauthor{\bsnm{Rockstr{\"{o}}m}, \binits{A.}},
\bauthor{\bsnm{Saracco}, \binits{R.}}:
\batitle{{SDL-CCITT} specification and description language}.
\bjtitle{IEEE Trans. Commun.}
\bvolume{30}(\bissue{6}),
\bfpage{1310}--\blpage{1318}
(\byear{1982})
\doiurl{10.1109/TCOM.1982.1095599}
\end{barticle}
\endbibitem

\bibitem[\protect\citeauthoryear{Bianchi et~al.}{2014}]{BianchiPV14}
\begin{barticle}
\bauthor{\bsnm{Bianchi}, \binits{A.}},
\bauthor{\bsnm{Pizzutilo}, \binits{S.}},
\bauthor{\bsnm{Vessio}, \binits{G.}}:
\batitle{Suitability of abstract state machines for discussing mobile ad-hoc
  networks}.
\bjtitle{Global Journal of Advanced Software Engineering}
\bvolume{1},
\bfpage{29}--\blpage{38}
(\byear{2014}).
\bcomment{\url{https://www.researchgate.net/publication/268813941_Suitability_of_Abstract_State_Machines_for_Discussing_Mobile_Ad-hoc_Networks}}
\end{barticle}
\endbibitem

\bibitem[\protect\citeauthoryear{Bianchi et~al.}{2017}]{BianchiPV17}
\begin{bchapter}
\bauthor{\bsnm{Bianchi}, \binits{A.}},
\bauthor{\bsnm{Pizzutilo}, \binits{S.}},
\bauthor{\bsnm{Vessio}, \binits{G.}}:
\bctitle{Intercepting blackhole attacks in {MANETs}: An {ASM}-based model}.
In: \bbtitle{Proc. 1st Workshop on Formal Approaches for Advanced Computing
  Systems (FAACS)}.
\bsertitle{LNCS},
vol. \bseriesno{10729},
pp. \bfpage{137}--\blpage{152}.
\bpublisher{Springer},
\blocation{Heidelberg}
(\byear{2017}).
\doiurl{10.1007/978-3-319-74781-1\_10}
\end{bchapter}
\endbibitem

\bibitem[\protect\citeauthoryear{Hoffmann and
  Mossakowski}{2002}]{HoffmannMossakowski02}
\begin{bchapter}
\bauthor{\bsnm{Hoffmann}, \binits{K.}},
\bauthor{\bsnm{Mossakowski}, \binits{T.}}:
\bctitle{Algebraic higher-order nets: Graphs and {Petri} nets as tokens}.
In: \bbtitle{Proc. 16th Workshop on Recent Trends in Algebraic Development
  Techniques (WADT)}.
\bsertitle{LNCS},
vol. \bseriesno{2755},
pp. \bfpage{253}--\blpage{267}.
\bpublisher{Springer},
\blocation{Heidelberg}
(\byear{2002}).
\doiurl{10.1007/978-3-540-40020-2\_14}
\end{bchapter}
\endbibitem

\bibitem[\protect\citeauthoryear{Bottoni et~al.}{2006}]{BottoniDHM06}
\begin{barticle}
\bauthor{\bsnm{Bottoni}, \binits{P.}},
\bauthor{\bsnm{{De Rosa}}, \binits{F.}},
\bauthor{\bsnm{Hoffmann}, \binits{K.}},
\bauthor{\bsnm{Mecella}, \binits{M.}}:
\batitle{Applying algebraic approaches for modeling workflows and their
  transformations in mobile networks}.
\bjtitle{Mob. Inf. Syst.}
\bvolume{2}(\bissue{1}),
\bfpage{51}--\blpage{76}
(\byear{2006})
\doiurl{10.1155/2006/704187}
\end{barticle}
\endbibitem

\bibitem[\protect\citeauthoryear{Padberg et~al.}{2007}]{PadbergHEMBE07}
\begin{bchapter}
\bauthor{\bsnm{Padberg}, \binits{J.}},
\bauthor{\bsnm{Hoffmann}, \binits{K.}},
\bauthor{\bsnm{Ehrig}, \binits{H.}},
\bauthor{\bsnm{Modica}, \binits{T.}},
\bauthor{\bsnm{Biermann}, \binits{E.}},
\bauthor{\bsnm{Ermel}, \binits{C.}}:
\bctitle{Maintaining consistency in layered architectures of mobile ad-hoc
  networks}.
In: \bbtitle{Proc. 10th Conference on Fundamental Approaches to Software
  Engineering (FASE)}.
\bsertitle{LNCS},
vol. \bseriesno{4422},
pp. \bfpage{383}--\blpage{397}.
\bpublisher{Springer},
\blocation{Heidelberg}
(\byear{2007}).
\doiurl{10.1007/978-3-540-71289-3\_29}
\end{bchapter}
\endbibitem

\bibitem[\protect\citeauthoryear{Biermann et~al.}{2008}]{BiermannHP08}
\begin{botherref}
\oauthor{\bsnm{Biermann}, \binits{E.}},
\oauthor{\bsnm{Hoffmann}, \binits{K.}},
\oauthor{\bsnm{Padberg}, \binits{J.}}:
Layered architecture consistency for {MANETs}: Introducing new team members.
Electron. Commun. Eur. Assoc. Softw. Sci. Technol.
\textbf{12}
(2008)
\doiurl{10.14279/tuj.eceasst.12.269}
\end{botherref}
\endbibitem

\bibitem[\protect\citeauthoryear{Archibald et~al.}{2021}]{ArchibaldKS21}
\begin{barticle}
\bauthor{\bsnm{Archibald}, \binits{B.}},
\bauthor{\bsnm{Kulcs\'ar}, \binits{G.}},
\bauthor{\bsnm{Sevegnani}, \binits{M.}}:
\batitle{A tale of two graph models: A case study in wireless sensor networks}.
\bjtitle{Form. Asp. Comp.}
\bvolume{33}(\bissue{6}),
\bfpage{1249}--\blpage{1277}
(\byear{2021})
\doiurl{10.1007/s00165-021-00558-z}
\end{barticle}
\endbibitem

\bibitem[\protect\citeauthoryear{Ehrig et~al.}{1973}]{EhrigPS73}
\begin{bchapter}
\bauthor{\bsnm{Ehrig}, \binits{H.}},
\bauthor{\bsnm{Pfender}, \binits{M.}},
\bauthor{\bsnm{Schneider}, \binits{H.J.}}:
\bctitle{Graph-grammars: An algebraic approach}.
In: \bbtitle{Proc. 14th Symposium on Switching and Automata Theory},
pp. \bfpage{167}--\blpage{180}.
\bpublisher{IEEE},
\blocation{New York}
(\byear{1973}).
\doiurl{10.1109/SWAT.1973.11}
\end{bchapter}
\endbibitem

\bibitem[\protect\citeauthoryear{Milner}{2001}]{Milner01}
\begin{bchapter}
\bauthor{\bsnm{Milner}, \binits{R.}}:
\bctitle{Bigraphical reactive systems}.
In: \bbtitle{Proc. 12th Conference on Concurrency Theory (CONCUR)}.
\bsertitle{LNCS},
vol. \bseriesno{2154},
pp. \bfpage{16}--\blpage{35}.
\bpublisher{Springer},
\blocation{Heidelberg}
(\byear{2001}).
\doiurl{10.1007/3-540-44685-0\_2}
\end{bchapter}
\endbibitem

\bibitem[\protect\citeauthoryear{Albalwe et~al.}{2024}]{AlbalweAS24}
\begin{barticle}
\bauthor{\bsnm{Albalwe}, \binits{M.}},
\bauthor{\bsnm{Archibald}, \binits{B.}},
\bauthor{\bsnm{Sevegnani}, \binits{M.}}:
\batitle{Modelling and analysing routing protocols diagrammatically with
  bigraphs}.
\bjtitle{Formal Aspects Comput.}
\bvolume{36}(\bissue{3}),
\bfpage{17}--\blpage{11725}
(\byear{2024})
\doiurl{10.1145/3685934}
\end{barticle}
\endbibitem

\bibitem[\protect\citeauthoryear{Alexander et~al.}{2012}]{AlexanderBVHPTLSKW12}
\begin{botherref}
\oauthor{\bsnm{Alexander}, \binits{R.}},
\oauthor{\bsnm{Brandt}, \binits{A.}},
\oauthor{\bsnm{Vasseur}, \binits{J.P.}},
\oauthor{\bsnm{Hui}, \binits{J.}},
\oauthor{\bsnm{Pister}, \binits{K.}},
\oauthor{\bsnm{Thubert}, \binits{P.}},
\oauthor{\bsnm{Levis}, \binits{P.}},
\oauthor{\bsnm{Struik}, \binits{R.}},
\oauthor{\bsnm{Kelsey}, \binits{R.}},
\oauthor{\bsnm{Winter}, \binits{T.}}:
RPL: IPv6 Routing Protocol for Low-power and LossyNetworks. RFC 6550.
IETF,
(2012).
IETF.
\url{https://www.rfc-editor.org/info/rfc6550}
\end{botherref}
\endbibitem

\bibitem[\protect\citeauthoryear{Camp et~al.}{2002}]{CampBD02}
\begin{barticle}
\bauthor{\bsnm{Camp}, \binits{T.}},
\bauthor{\bsnm{Boleng}, \binits{J.}},
\bauthor{\bsnm{Davies}, \binits{V.}}:
\batitle{A survey of mobility models for ad hoc network research}.
\bjtitle{Wirel. Commun. Mob. Comput.}
\bvolume{2},
\bfpage{483}--\blpage{502}
(\byear{2002})
\doiurl{10.1002/wcm.72}
\end{barticle}
\endbibitem

\bibitem[\protect\citeauthoryear{Patsouris}{2001}]{Patsouris01}
\begin{barticle}
\bauthor{\bsnm{Patsouris}, \binits{P.A.}}:
\batitle{Algebraic modeling of an ad hoc network for mobile computing}.
\bjtitle{J. Parallel Distributed Comput.}
\bvolume{61}(\bissue{7}),
\bfpage{884}--\blpage{897}
(\byear{2001})
\doiurl{10.1006/jpdc.2000.1718}
\end{barticle}
\endbibitem

\bibitem[\protect\citeauthoryear{Romanowska and
  Smith}{1983}]{RomanowskaSmith83}
\begin{bbook}
\bauthor{\bsnm{Romanowska}, \binits{A.B.}},
\bauthor{\bsnm{Smith}, \binits{J.D.H.}}:
\bbtitle{Modal Theory: An Algebraic Approach to Order, Geometry, and
  Convexity}.
\bpublisher{Heldermann Verlag},
\blocation{Berlin}
(\byear{1983})
\end{bbook}
\endbibitem

\bibitem[\protect\citeauthoryear{H{\"{o}}fner and
  McIver}{2014}]{HofnerMcIver14}
\begin{barticle}
\bauthor{\bsnm{H{\"{o}}fner}, \binits{P.}},
\bauthor{\bsnm{McIver}, \binits{A.}}:
\batitle{Hopscotch - {R}eaching the target hop by hop}.
\bjtitle{J. Log. Algebraic Methods Program.}
\bvolume{83}(\bissue{2}),
\bfpage{212}--\blpage{224}
(\byear{2014})
\doiurl{10.1016/j.jlap.2014.02.009}
\end{barticle}
\endbibitem

\bibitem[\protect\citeauthoryear{Godskesen and Nanz}{2009}]{GodskesenNanz09}
\begin{bchapter}
\bauthor{\bsnm{Godskesen}, \binits{J.C.}},
\bauthor{\bsnm{Nanz}, \binits{S.}}:
\bctitle{Mobility models and behavioural equivalence for wireless networks}.
In: \bbtitle{Proc. 11th Conference on Coordination Models and Languages
  (COORDINATION)}.
\bsertitle{LNCS},
vol. \bseriesno{5521},
pp. \bfpage{106}--\blpage{122}.
\bpublisher{Springer},
\blocation{Heidelberg}
(\byear{2009}).
\doiurl{10.1007/978-3-642-02053-7\_6}
\end{bchapter}
\endbibitem

\bibitem[\protect\citeauthoryear{Wu et~al.}{2014}]{WuLZZ14}
\begin{bchapter}
\bauthor{\bsnm{Wu}, \binits{X.}},
\bauthor{\bsnm{Liu}, \binits{S.}},
\bauthor{\bsnm{Zhu}, \binits{H.}},
\bauthor{\bsnm{Zhao}, \binits{Y.}}:
\bctitle{Reasoning about group-based mobility in {MANETs}}.
In: \bbtitle{Proc. 20th Pacific Rim Symposium on Dependable Computing, (PRDC)},
pp. \bfpage{244}--\blpage{253}.
\bpublisher{IEEE},
\blocation{New York}
(\byear{2014}).
\doiurl{10.1109/PRDC.2014.39}
\end{bchapter}
\endbibitem

\bibitem[\protect\citeauthoryear{Pei et~al.}{1999}]{PeiGHC99}
\begin{bchapter}
\bauthor{\bsnm{Pei}, \binits{G.}},
\bauthor{\bsnm{Gerla}, \binits{M.}},
\bauthor{\bsnm{Hong}, \binits{X.}},
\bauthor{\bsnm{Chiang}, \binits{C.}}:
\bctitle{A wireless hierarchical routing protocol with group mobility}.
In: \bbtitle{Proc. Wireless Communications and Networking Conference, (WCNC)},
pp. \bfpage{1538}--\blpage{1542}.
\bpublisher{IEEE},
\blocation{New York}
(\byear{1999}).
\doiurl{10.1109/WCNC.1999.796996}
\end{bchapter}
\endbibitem

\bibitem[\protect\citeauthoryear{Song and Godskesen}{2010}]{SongGodskesen10}
\begin{bchapter}
\bauthor{\bsnm{Song}, \binits{L.}},
\bauthor{\bsnm{Godskesen}, \binits{J.C.}}:
\bctitle{Probabilistic mobility models for mobile and wireless networks}.
In: \bbtitle{Proc. 6th IFIP WG 2.2 Conference on Theoretical Computer Science
  (IFIP TCS)}.
\bsertitle{{IFIP} Advances in Information and Communication Technology},
vol. \bseriesno{323},
pp. \bfpage{86}--\blpage{100}.
\bpublisher{Springer},
\blocation{Heidelberg}
(\byear{2010}).
\doiurl{10.1007/978-3-642-15240-5\_7}
\end{bchapter}
\endbibitem

\bibitem[\protect\citeauthoryear{Cheshire et~al.}{2005}]{CheshireAG05}
\begin{botherref}
\oauthor{\bsnm{Cheshire}, \binits{S.}},
\oauthor{\bsnm{Aboba}, \binits{B.}},
\oauthor{\bsnm{Guttman}, \binits{E.}}:
Dynamic Configuration of IPv4 Link-Local Addresses. RFC 3927.
IETF,
(2005).
IETF.
\url{https://datatracker.ietf.org/doc/html/rfc3927}
\end{botherref}
\endbibitem

\bibitem[\protect\citeauthoryear{Song and Godskesen}{2012}]{SongGodskesen12}
\begin{bchapter}
\bauthor{\bsnm{Song}, \binits{L.}},
\bauthor{\bsnm{Godskesen}, \binits{J.C.}}:
\bctitle{Broadcast abstraction in a stochastic calculus for mobile networks}.
In: \bbtitle{Proc. 7th IFIP WG 2.2 Conference on Theoretical Computer Science
  (IFIP TCS)}.
\bsertitle{LNCS},
vol. \bseriesno{7604},
pp. \bfpage{342}--\blpage{356}.
\bpublisher{Springer},
\blocation{Heidelberg}
(\byear{2012}).
\doiurl{10.1007/978-3-642-33475-7\_24}
\end{bchapter}
\endbibitem

\bibitem[\protect\citeauthoryear{Ghassemi et~al.}{2010}]{GhassemiMF10}
\begin{bchapter}
\bauthor{\bsnm{Ghassemi}, \binits{F.}},
\bauthor{\bsnm{Movaghar}, \binits{A.}},
\bauthor{\bsnm{Fokkink}, \binits{W.J.}}:
\bctitle{Towards performance evaluation of mobile ad hoc network protocols}.
In: \bbtitle{Proc. 10th Conference on Application of Concurrency to System
  Design (ACSD)},
pp. \bfpage{85}--\blpage{92}.
\bpublisher{IEEE},
\blocation{New York}
(\byear{2010}).
\doiurl{10.1109/ACSD.2010.20}
\end{bchapter}
\endbibitem

\bibitem[\protect\citeauthoryear{Hinton et~al.}{2006}]{HintonKNP06}
\begin{bchapter}
\bauthor{\bsnm{Hinton}, \binits{A.}},
\bauthor{\bsnm{Kwiatkowska}, \binits{M.Z.}},
\bauthor{\bsnm{Norman}, \binits{G.}},
\bauthor{\bsnm{Parker}, \binits{D.}}:
\bctitle{{PRISM}: A tool for automatic verification of probabilistic systems}.
In: \bbtitle{Proc. 12th Conference on Tools and Algorithms for the Construction
  and Analysis of Systems (TACAS)}.
\bsertitle{LNCS},
vol. \bseriesno{3920},
pp. \bfpage{441}--\blpage{444}.
\bpublisher{Springer},
\blocation{Heidelberg}
(\byear{2006}).
\doiurl{10.1007/11691372\_29}
\end{bchapter}
\endbibitem

\bibitem[\protect\citeauthoryear{Mouradian and
  Aug{\'{e}}{-}Blum}{2013}]{MouradianAuge-Blum13}
\begin{bchapter}
\bauthor{\bsnm{Mouradian}, \binits{A.}},
\bauthor{\bsnm{Aug{\'{e}}{-}Blum}, \binits{I.}}:
\bctitle{Formal verification of real-time wireless sensor networks protocols
  with realistic radio links}.
In: \bbtitle{Proc. 21st Conference on Real-Time Networks and Systems (RTNS)},
pp. \bfpage{213}--\blpage{222}.
\bpublisher{ACM},
\blocation{New York}
(\byear{2013}).
\doiurl{10.1145/2516821.2516833}
\end{bchapter}
\endbibitem

\bibitem[\protect\citeauthoryear{Roedig et~al.}{2006}]{RoedigBS06}
\begin{bchapter}
\bauthor{\bsnm{Roedig}, \binits{U.}},
\bauthor{\bsnm{Barroso}, \binits{A.M.}},
\bauthor{\bsnm{Sreenan}, \binits{C.J.}}:
\bctitle{{f-MAC}: A deterministic media access control protocol without time
  synchronization}.
In: \bbtitle{Proc. 3rd European Workshop on Wireless Sensor Networks (EWSN)}.
\bsertitle{LNCS},
vol. \bseriesno{3868},
pp. \bfpage{276}--\blpage{291}.
\bpublisher{Springer},
\blocation{Heidelberg}
(\byear{2006}).
\doiurl{10.1007/11669463\_21}
\end{bchapter}
\endbibitem

\bibitem[\protect\citeauthoryear{Kamali and Katoen}{2020}]{KamaliKatoen20}
\begin{bchapter}
\bauthor{\bsnm{Kamali}, \binits{M.}},
\bauthor{\bsnm{Katoen}, \binits{J.}}:
\bctitle{Probabilistic model checking of {AODV}}.
In: \bbtitle{Proc. 17th Conference on Quantitative Evaluation of Systems
  (QEST)}.
\bsertitle{LNCS},
vol. \bseriesno{12289},
pp. \bfpage{54}--\blpage{73}.
\bpublisher{Springer},
\blocation{Heidelberg}
(\byear{2020}).
\doiurl{10.1007/978-3-030-59854-9\_6}
\end{bchapter}
\endbibitem

\bibitem[\protect\citeauthoryear{Bohnenkamp et~al.}{2006}]{BohnenkampDHK06}
\begin{barticle}
\bauthor{\bsnm{Bohnenkamp}, \binits{H.C.}},
\bauthor{\bsnm{D'Argenio}, \binits{P.R.}},
\bauthor{\bsnm{Hermanns}, \binits{H.}},
\bauthor{\bsnm{Katoen}, \binits{J.}}:
\batitle{{MODEST}: A compositional modeling formalism for hard and softly timed
  systems}.
\bjtitle{IEEE Trans. Software Eng.}
\bvolume{32}(\bissue{10}),
\bfpage{812}--\blpage{830}
(\byear{2006})
\doiurl{10.1109/TSE.2006.104}
\end{barticle}
\endbibitem

\bibitem[\protect\citeauthoryear{Fehnker et~al.}{2013}]{FehnkerHKM13}
\begin{bchapter}
\bauthor{\bsnm{Fehnker}, \binits{A.}},
\bauthor{\bsnm{H{\"{o}}fner}, \binits{P.}},
\bauthor{\bsnm{Kamali}, \binits{M.}},
\bauthor{\bsnm{Mehta}, \binits{V.}}:
\bctitle{Topology-based mobility models for wireless networks}.
In: \bbtitle{Proc. 10th Conference on Quantitative Evaluation of Systems
  (QEST)}.
\bsertitle{LNCS},
vol. \bseriesno{8054},
pp. \bfpage{389}--\blpage{404}.
\bpublisher{Springer},
\blocation{Heidelberg}
(\byear{2013}).
\doiurl{10.1007/978-3-642-40196-1\_32}
\end{bchapter}
\endbibitem

\bibitem[\protect\citeauthoryear{{van Hoesel} and
  Havinga}{2004}]{HoeselHavinga04}
\begin{bchapter}
\bauthor{\bsnm{{van Hoesel}}, \binits{L.F.W.}},
\bauthor{\bsnm{Havinga}, \binits{P.J.M.}}:
\bctitle{A lightweight medium access protocol ({LMAC}) for wireless sensor
  networks: Reducing preamble transmissions and transceiver state switches}.
In: \bbtitle{Proc. 1st Workshop on Networked Sensing Systems (INSS)},
pp. \bfpage{205}--\blpage{208}
(\byear{2004}).
\burl{https://www.researchgate.net/publication/242150051_A_Lightweight_Medium_Access_Protocol_LMAC_for_Wireless_Sensor_Networks_Reducing_Preamble_Transmissions_and_Transceiver_State_Switches}
\end{bchapter}
\endbibitem

\bibitem[\protect\citeauthoryear{Wu and Zhu}{2015}]{WuZhu15}
\begin{bchapter}
\bauthor{\bsnm{Wu}, \binits{X.}},
\bauthor{\bsnm{Zhu}, \binits{H.}}:
\bctitle{A calculus for wireless sensor networks from quality perspective}.
In: \bbtitle{Proc. 16th Symposium on High Assurance Systems Engineering
  (HASE)},
pp. \bfpage{223}--\blpage{231}.
\bpublisher{IEEE},
\blocation{New York}
(\byear{2015}).
\doiurl{10.1109/HASE.2015.40}
\end{bchapter}
\endbibitem

\bibitem[\protect\citeauthoryear{Wu et~al.}{2016}]{WuZZ16}
\begin{bchapter}
\bauthor{\bsnm{Wu}, \binits{X.}},
\bauthor{\bsnm{Zhao}, \binits{Y.}},
\bauthor{\bsnm{Zhu}, \binits{H.}}:
\bctitle{Integrating a calculus with mobility and quality for wireless sensor
  networks}.
In: \bbtitle{Proc. 17th Symposium on High Assurance Systems Engineering
  (HASE)},
pp. \bfpage{220}--\blpage{227}.
\bpublisher{IEEE},
\blocation{New York}
(\byear{2016}).
\doiurl{10.1109/HASE.2016.29}
\end{bchapter}
\endbibitem

\bibitem[\protect\citeauthoryear{Wu et~al.}{2019}]{WuZX19}
\begin{bchapter}
\bauthor{\bsnm{Wu}, \binits{X.}},
\bauthor{\bsnm{Zhu}, \binits{H.}},
\bauthor{\bsnm{Xie}, \binits{W.}}:
\bctitle{{UTP} semantics of a calculus for mobile ad hoc networks}.
In: \bbtitle{Proc. 7th Symposium on Unifying Theories of Programming (UTP)}.
\bsertitle{LNCS},
vol. \bseriesno{11885},
pp. \bfpage{198}--\blpage{216}.
\bpublisher{Springer},
\blocation{Heidelberg}
(\byear{2019}).
\doiurl{10.1007/978-3-030-31038-7\_10}
\end{bchapter}
\endbibitem

\bibitem[\protect\citeauthoryear{Xie et~al.}{2019}]{XieZWV19}
\begin{barticle}
\bauthor{\bsnm{Xie}, \binits{W.}},
\bauthor{\bsnm{Zhu}, \binits{H.}},
\bauthor{\bsnm{Wu}, \binits{X.}},
\bauthor{\bsnm{Vinh}, \binits{P.C.}}:
\batitle{Formal verification of {mCWQ} using extended {Hoare} logic}.
\bjtitle{Mob. Networks Appl.}
\bvolume{24}(\bissue{1}),
\bfpage{134}--\blpage{144}
(\byear{2019})
\doiurl{10.1007/s11036-018-1142-8}
\end{barticle}
\endbibitem

\bibitem[\protect\citeauthoryear{Hoare}{1969}]{Hoare69}
\begin{barticle}
\bauthor{\bsnm{Hoare}, \binits{C.A.R.}}:
\batitle{An axiomatic basis for computer programming}.
\bjtitle{Commun. ACM}
\bvolume{12}(\bissue{10}),
\bfpage{576}--\blpage{580}
(\byear{1969})
\doiurl{10.1145/363235.363259}
\end{barticle}
\endbibitem

\bibitem[\protect\citeauthoryear{Liu et~al.}{2016}]{LiuOM16}
\begin{barticle}
\bauthor{\bsnm{Liu}, \binits{S.}},
\bauthor{\bsnm{{\"{O}}lveczky}, \binits{P.C.}},
\bauthor{\bsnm{Meseguer}, \binits{J.}}:
\batitle{Modeling and analyzing mobile ad hoc networks in {Real-Time Maude}}.
\bjtitle{J. Log. Algebraic Methods Program.}
\bvolume{85}(\bissue{1}),
\bfpage{34}--\blpage{66}
(\byear{2016})
\doiurl{10.1016/j.jlamp.2015.05.002}
\end{barticle}
\endbibitem

\bibitem[\protect\citeauthoryear{{\"{O}}lveczky and
  Meseguer}{2000}]{OlveczkyMeseguer00}
\begin{bchapter}
\bauthor{\bsnm{{\"{O}}lveczky}, \binits{P.C.}},
\bauthor{\bsnm{Meseguer}, \binits{J.}}:
\bctitle{Real-{T}ime {Maude}: {A} tool for simulating and analyzing real-time
  and hybrid systems}.
In: \bbtitle{Proc. 3rd Workshop on Rewriting Logic and Its Applications
  (WRLA)}.
\bsertitle{ENTCS},
vol. \bseriesno{36},
pp. \bfpage{361}--\blpage{382}.
\bpublisher{Elsevier},
\blocation{Amsterdam}
(\byear{2000}).
\doiurl{10.1016/S1571-0661(05)80134-3}
\end{bchapter}
\endbibitem

\bibitem[\protect\citeauthoryear{Lepri et~al.}{2015}]{LepriAO15}
\begin{barticle}
\bauthor{\bsnm{Lepri}, \binits{D.}},
\bauthor{\bsnm{{\'{A}}brah{\'{a}}m}, \binits{E.}},
\bauthor{\bsnm{{\"{O}}lveczky}, \binits{P.C.}}:
\batitle{Sound and complete timed {CTL} model checking of timed {Kripke}
  structures and real-time rewrite theories}.
\bjtitle{Sci. Comput. Program.}
\bvolume{99},
\bfpage{128}--\blpage{192}
(\byear{2015})
\doiurl{10.1016/j.scico.2014.06.006}
\end{barticle}
\endbibitem

\bibitem[\protect\citeauthoryear{Gallina et~al.}{2016}]{GallinaMR16}
\begin{barticle}
\bauthor{\bsnm{Gallina}, \binits{L.}},
\bauthor{\bsnm{Marin}, \binits{A.}},
\bauthor{\bsnm{Rossi}, \binits{S.}}:
\batitle{Connectivity and energy-aware preorders for mobile ad-hoc networks}.
\bjtitle{Telecommun. Syst.}
\bvolume{63}(\bissue{2}),
\bfpage{307}--\blpage{333}
(\byear{2016})
\doiurl{10.1007/s11235-015-0122-6}
\end{barticle}
\endbibitem

\bibitem[\protect\citeauthoryear{Gallina et~al.}{2013}]{GallinaMRHK13}
\begin{bchapter}
\bauthor{\bsnm{Gallina}, \binits{L.}},
\bauthor{\bsnm{Marin}, \binits{A.}},
\bauthor{\bsnm{Rossi}, \binits{S.}},
\bauthor{\bsnm{Han}, \binits{T.}},
\bauthor{\bsnm{Kwiatkowska}, \binits{M.Z.}}:
\bctitle{A process algebraic framework for estimating the energy consumption in
  ad-hoc wireless sensor networks}.
In: \bbtitle{Proc. 16th Conference on Modeling, Analysis and Simulation of
  Wireless and Mobile Systems (MSWiM)},
pp. \bfpage{255}--\blpage{262}.
\bpublisher{ACM},
\blocation{New York}
(\byear{2013}).
\doiurl{10.1145/2507924.2507958}
\end{bchapter}
\endbibitem

\bibitem[\protect\citeauthoryear{Haas et~al.}{2006}]{HaasHL06}
\begin{barticle}
\bauthor{\bsnm{Haas}, \binits{Z.J.}},
\bauthor{\bsnm{Halpern}, \binits{J.Y.}},
\bauthor{\bsnm{Li}, \binits{L.}}:
\batitle{Gossip-based ad hoc routing}.
\bjtitle{{IEEE/ACM} Trans. Netw.}
\bvolume{14}(\bissue{3}),
\bfpage{479}--\blpage{491}
(\byear{2006})
\doiurl{10.1109/TNET.2006.876186}
\end{barticle}
\endbibitem

\bibitem[\protect\citeauthoryear{Bugliesi et~al.}{2014}]{BugliesiGHMR14}
\begin{barticle}
\bauthor{\bsnm{Bugliesi}, \binits{M.}},
\bauthor{\bsnm{Gallina}, \binits{L.}},
\bauthor{\bsnm{Hamadou}, \binits{S.}},
\bauthor{\bsnm{Marin}, \binits{A.}},
\bauthor{\bsnm{Rossi}, \binits{S.}}:
\batitle{Behavioural equivalences and interference metrics for mobile ad-hoc
  networks}.
\bjtitle{Perform. Evaluation}
\bvolume{73},
\bfpage{41}--\blpage{72}
(\byear{2014})
\doiurl{10.1016/j.peva.2013.11.003}
\end{barticle}
\endbibitem

\end{thebibliography}

\end{document}